\documentstyle[pra,aps,eqsecnum,amsfonts]{revtex}

\begin{document}

\draft

\title{Geometric Phase, Curvature, and Extrapotentials in Constrained Quantum Systems}

\author{Kevin A. Mitchell}

\address{Department of Physics, University of California, 
Berkeley, California 94720}

\date{\today}

\maketitle

\begin{abstract}
We derive an effective Hamiltonian for a quantum system constrained to
a submanifold (the constraint manifold) of configuration space (the
ambient space) by an infinite restoring force.  We pay special
attention to how this Hamiltonian depends on quantities which are
external to the constraint manifold, such as the external curvature of
the constraint manifold, the (Riemannian) curvature of the ambient
space, and the constraining potential.  In particular, we find the
remarkable fact that the twisting of the constraining potential
appears as a gauge potential in the constrained Hamiltonian.  This gauge
potential is an example of geometric phase, closely related to that
originally discussed by Berry.  The constrained Hamiltonian also
contains an effective potential depending on the external curvature of
the constraint manifold, the curvature of the ambient space, and the
twisting of the constraining potential.  The general nature of our
analysis allows applications to a wide variety of problems, such as
rigid molecules, the evolution of molecular systems along reaction
paths, and quantum strip waveguides.
\end{abstract}

\pacs{03.65.Bz,02.40.Ky,33.20.Vq,34.10.+x}

\section{Introduction}

We derive an effective Hamiltonian for a quantum system subject to an
infinite restoring force.  Though our results are quite general, we
are motivated by several specific applications, such as stiff
molecular bonds in rigid molecules and clusters of rigid molecules,
molecular systems evolving along reaction paths, and electrons
confined to quantum strip waveguides.

For comparison, consider first a classical system.  We have in mind a
system which initially occupies any position in the full configuration
space (called the ambient space) but is subsequently confined to a
submanifold (called the constraint manifold) by the introduction of a
restoring force, which in a certain limit becomes infinite.  Here, the
Hamiltonian is simply the kinetic energy plus the constraining
potential, which we assume is constant along the constraint manifold.
Assuming the initial velocity is tangent to the constraint manifold,
it is well known that the system's trajectory remains on the
constraint manifold and that its motion is determined solely by the
form of the kinetic energy tangent to the manifold
\cite{Koppe71,vanKampen85}.  This kinetic energy, in turn, depends
only on the Riemannian metric of the constraint manifold.  Thus, the
motion of the constrained classical system depends only on the
internal metric of the constraint manifold and is independent of the
ambient space, the embedding of the constraint manifold within the
ambient space, or the details of the constraining potential.  It is a
remarkable fact, then, that for a quantum system this is no longer
true.  The constrained quantum Hamiltonian depends on the curvature of
the ambient space, the external curvatures of the constraint manifold,
and on properties of the constraining potential.

It has already been known for some time that a constrained quantum system
``senses'' the local neighborhood of the constraint manifold
\cite{Marcus66,Jensen71,Switkes77,daCosta81,daCosta82,daCosta83,vanKampen84,daCosta86,Kaplan97}.
As a simple example, consider a quantum system whose motion transverse
to the constraint manifold is in the ground state.  Due to the
conservation of the transverse action, the constrained quantum system
sees the transverse zero point energy as an effective potential on the
constraint manifold.  (We call this the adiabatic potential.)  The
adiabatic potential also appears classically if the initial velocity
of the system has a nonzero component normal to the constraint
manifold.  Classically, the adiabatic potential can always be
eliminated by choosing an initial velocity tangent to the constraint
manifold.  Quantum mechanically, however, due to the Heisenberg
principle, the transverse action and hence the adiabatic potential can
never be eliminated.

The present paper focuses on effects of the ambient space and
constraining potential other than the adiabatic potential.  Following
da Costa\cite{daCosta81,daCosta82,daCosta83,daCosta86}, we assume that
the constraining potential has the same form at each point of the
constraint manifold.  The adiabatic potential is therefore constant
along the constraint manifold and can subsequently be ignored.  (In
Sect.~\ref{s22}, we discuss briefly how a small amount of variation in
the adiabatic potential can be accommodated.)  In two noteworthy
papers\cite{daCosta81,daCosta82}, da Costa, using this assumption,
derived the effective Hamiltonian for a system of $n$ constrained
point particles.  This Hamiltonian contains two terms.  The first is
proportional to the Laplacian on the constraint manifold, and
therefore depends only on the internal metric of the constraint
manifold.  The second, however, is an effective potential, called the
extrapotential, which depends not only on the internal curvature, but
also the external curvature of the constraint manifold.  This
extrapotential is of order $\hbar^2$ and therefore vanishes in the
classical (and semi-classical) limit.  As a simple, yet illustrative,
example, consider a system defined on ${\Bbb R}^3$ constrained to lie
on a curve.  For this system, the extrapotential is $-\hbar^2/(8
\rho^2)$, where $\rho$ is the  radius of curvature.  This
result was obtained by da Costa \cite{daCosta81}; the same result was
obtained earlier by Marcus
\cite{Marcus66} and Switkes et al. \cite{Switkes77} for curves in
${\Bbb R}^2$.  Others have also studied this order $\hbar^2$
extrapotential, including Jensen and Koppe \cite{Jensen71} and Kaplan,
Maitra, and Heller \cite{Kaplan97}.  Since the extrapotential depends
on the external curvature, it can never be derived from a procedure
which quantizes the constrained classical system (which depends only
on the internal curvature), an approach common in the literature of
constrained quantum mechanics. (See, for example, the review of DeWitt
\cite{DeWitt57}.)

As mentioned above, once the constraining potential is defined at one
point of the constraint manifold, the constraining potential at all
other points must have the same form.  However, this requirement does
not completely determine the constraining potential at all points
since the orientation of the potential is left unspecified.  In other
words, the equipotentials surrounding the constraint manifold can
twist in some unspecified manner as one moves along the manifold.

Da Costa fixed the twisting ambiguity by imposing what we call a ``no
twist'' condition on the potential.  Physically, this condition
requires the restoring forces in the neighborhood of the constraint
manifold to be normal to the manifold.  It can be viewed as an
extension of the fact that in classical mechanics nondissipative
forces are normal to the constraint manifold at the point of the
manifold itself.  Da Costa astutely realized that if the no twist
condition were violated, the motion transverse to the constraint
manifold would be coupled to the motion tangent to the manifold and
the Schr\"{o}dinger equation would not separate.

For some submanifolds there exist no potentials which satisfy the no
twist condition.  In Ref.~\cite{daCosta82} da Costa derived a local
geometric criterion on the external curvature of a submanifold which
was necessary and sufficient to determine the existence of a
non-twisting potential.  Unfortunately, several common examples of
constrained systems do not satisfy this criterion.  For example,
consider a model of a polymer by $n>2$ point particles where the
distances between each particle $i$ and its neighbor $i + 1$ are
fixed.  (This model also applies to the double pendulum.)  These
systems all fail the criterion
\cite{Alvarez-Estrada92} as does a system of $n>2$ point particles
constrained to form a rigid body.  Even if a given submanifold can
have a non-twisting constraining potential, there is no guarantee that
this potential is the one dictated by the physics of the system under
consideration.

The principal objective of this paper is to derive an effective
Hamiltonian for a constrained quantum system with arbitrary twisting of the potential.
The presence of the potential twist leads to several qualitative
changes in the Hamiltonian.  First, the Hamiltonian is no longer a
scalar operator, but rather a $k \times k$ matrix of operators acting
on a $k$-dimensional vector-valued wave function defined over the
constraint manifold.  Here, $k$ is the degeneracy of the transverse
modes with each component of the vector wave function representing a
different transverse mode.  Of course, if one chooses a nondegenerate
transverse mode, the Hamiltonian reduces to a single component.

Perhaps the most interesting consequence of dropping the no twist
condition is the emergence of a $U(k)$ gauge potential, or connection,
in the constrained Hamiltonian.  This gauge potential is a coupling
between the twisting of the potential and the generalized angular
momentum of the transverse modes.  Modes with no such angular momentum
are unaffected by the potential twist.  This gauge potential is an
example of geometric phase and is closely related to the phase
originally introduced by Berry in the context of adiabatic transport
of quantum states \cite{Berry84}.  It is interesting to note that the
gauge potential is of order $\hbar^0$ and therefore, like the adiabatic
potential, is essentially of classical origin.

In addition to creating the gauge potential, the potential twist adds
a term to the extrapotential.  Unlike the extrapotential terms derived by da
Costa, the potential twist term is not a scalar function, but a $k
\times k$ matrix of such functions with possible off-diagonal terms
coupling  the degenerate transverse modes.  The potential twist term
depends on the standard deviation of the angular momentum of the
transverse modes and thus disappears for angular momentum eigenstates.

In some applications, the ambient space may not be flat.  For example,
the internal space of a molecule with $n>2$ atoms is not flat
\cite{Littlejohn97}.  Constraining such a molecule to a reaction path
therefore requires an analysis accounting for the ambient curvature.
We therefore do not assume in this paper that the ambient space is
flat. The effects of the ambient curvature are most notable as
additional terms in the extrapotential, although it also modifies the
curvature of the gauge potential.

This paper has the following organization.  In Sect.~\ref{s7}, we
introduce many of the key concepts by a simple example: that of a
system on ${\Bbb R}^3$ constrained to a curve.  Section~\ref{s7} is
purely expository, containing no derivations.  Section~\ref{s14}
briefly introduces the general problem.  In Sect.~\ref{s15} we focus
on the constraining potential.  We take care to precisely define the
requirement that it have the same form at all points of the constraint
manifold.  We also define a tensor which measures the twisting of the
potential.  In Sect.~\ref{s10} we specify how the potential is to
scale in $\epsilon$, where $\epsilon \rightarrow 0$ represents an
infinite constraining force.  The main computations of the paper are
in Sect.~\ref{s11} in which we expand the kinetic energy in $\epsilon$
and arrive at a preliminary expression, Eq.~(\ref{r32}), for the
constrained kinetic energy.  Section~\ref{s16} is devoted to deriving
various expressions for the extrapotential.  In Sect.~\ref{s3}, we
apply first order perturbation theory to transform to a set of
degenerate transverse modes, thereby obtaining Eqs.~(\ref{r105}) --
(\ref{r51}), which are the main results of the paper.
Section~\ref{s22} briefly discusses nonconstant constraining
potentials.  In Sect.~\ref{s13}, we study the geometric origins of the
gauge potential and various related connections.  We also compute
their curvatures.  Section~\ref{s17} contains some special cases,
including constraint manifolds of codimensions one and two,
rotationally invariant constraining potentials, and harmonic
constraining potentials.  In Sect.~\ref{s23}, we show that the gauge
potential vanishes for certain systems with reflection symmetry.
Conclusions are in Sect.~\ref{s18}.  There
are three Appendices.  Appendix~\ref{s8} contains a very brief review
of curves in ${\Bbb R}^3$.  Appendix~\ref{s1} is a review of the
second fundamental form.  Appendix~\ref{s9} summarizes an expression
we will need for the quantum kinetic energy.

\section{A Simple Example: A Curve in ${\Bbb R}^3$}

\label{s7}

The ultimate objective of this paper is to constrain a quantum wave
function, defined on an arbitrary manifold (the ambient space),
to a (locally) arbitrary submanifold (the constraint manifold)
via some general constraining potential.  Before solving the full
problem, however, it is instructive to consider a simple (though
certainly non-trivial), concrete example of the constraining
procedure.  We present no derivations here; our results will be
justified later in Sect.~\ref{s12}. 

We consider a curve embedded in flat three-dimensional space ${\Bbb
R}^3$ and parameterized by its arclength $\alpha$.  (See
Fig.~\ref{f1}.)  Such a curve is characterized by its curvature
$\kappa$ and torsion $\tau$.  (See Appendix~\ref{s8}.)  We take this
curve to be the axis of a quantum waveguide.  That is, there is a tube
enclosing the curve such that the potential is zero inside the tube
and infinite outside.  We assume the cross-section of the tube is
constant along the curve.  More precisely, if we cut the tube along a
plane normal to the curve (called hereafter a normal plane), the
cross-sectional shape of the tube is independent of where along the
curve we cut.  Two such cross-sections have the same shape if one can
be rotated into the other.  This rotational freedom permits the
cross-sectional shape to twist as one moves along the curve, even if
the curve itself is straight.  The orientation of the cross-sectional
shape is specified by two orthonormal vectors ${\bf E}_1$ and ${\bf
E}_2$ chosen from each normal plane along the curve.  The choice of
orthonormal frame $({\bf E}_1, {\bf E}_2)$ is such that the
cross-sectional shape (with respect to this frame) is independent of
$\alpha$.  In Fig.~\ref{f1}, the cross-section is a triangle with no
reflection symmetry.  Such symmetry is nongeneric and can cause
certain terms discussed below to vanish.  (See Sect.~\ref{s23}.)

We assume that the transverse dimensions of the tube are small
compared to the radius of curvature $\rho = \kappa^{-1}$ and the
inverse torsion $\tau^{-1}$.  We can then separate out the ``fast''
transverse degrees of freedom and obtain an effective one-dimensional
Hamiltonian in the ``slow'' longitudinal, or tangential, coordinate
$\alpha$.  To accomplish this separation, we pick a transverse mode
$\chi(u^1,u^2)$ of the waveguide.  Here $(u^1,u^2)$ are the Cartesian
coordinates in the normal plane taken with respect to the frame $({\bf
E}_1,{\bf E}_2)$; the quantities $(u^1,u^2, \alpha)$ thus coordinatize
the tube.  The transverse mode $\chi(u^1,u^2)$ is a normalized
eigenfunction of the transverse Hamiltonian $H_\perp = (\pi_1^2 +
\pi_2^2)/2 + V_\perp(u^1,u^2)$, where $\pi_j = -i \hbar \partial /
\partial u^j$, $j = 1,2$, and $V_\perp(u^1,u^2)$ is the potential
energy which defines the tube.  The eigenvalue of $H_\perp$
corresponding to $\chi$ is called the transverse energy.  For
simplicity, we assume that the transverse energy is nondegenerate.

To lowest order in the width of the tube, an eigenfunction $\psi$ of
the wave guide has the form

\begin{equation}
\psi(u^1,u^2,\alpha) 
= \chi(u^1,u^2) \phi(\alpha).
\end{equation}
As we take the limit where the transverse dimensions of the waveguide
shrink to zero (keeping the quantum numbers of the transverse mode
fixed), the transverse energy obviously tends toward infinity.
However, due to the constancy of the cross-sectional shape, this
transverse energy, though very large, is itself constant along the
curve.  We thus subtract it off, leaving a residual Hamiltonian
$H_\parallel$, which we call the constrained Hamiltonian.  The
constrained Hamiltonian acts only on $\phi$, resulting in the
Schr\"{o}dinger equation

\begin{equation}
H_\parallel \phi = E_\parallel \phi.
\end{equation}
The principal objective of this paper is to determine the form of this
constrained Hamiltonian.

As we will show later, the constrained Hamiltonian is not simply
$\pi_\parallel^2/2$ where $\pi_\parallel = -i \hbar \partial /
\partial \alpha$.  Rather, there are effects from the curvature
$\kappa$ and the rate at which the cross-sectional shape twists along
the curve.  To make this latter concept more precise, we define the
potential twist $S = {\bf E}_1 \cdot (d {\bf E}_2 / d \alpha)$ which
measures the rotation rate of the cross-sectional shape.  The
potential twist admits the following description.  Let $\theta$ denote
the angle between the principal normal $\hat{\bf n}$ (see Appendix
\ref{s8}) and the frame $({\bf E}_1,{\bf E}_2)$, specifically, $\hat{\bf n} \cdot
{\bf E}_1 = \cos \theta$, $\hat{\bf n} \cdot {\bf E}_2 = -\sin
\theta$.  Let $\omega = d \theta / d \alpha$ denote the rotation
rate of the frame $({\bf E}_1, {\bf E}_2)$ with respect to $\hat{\bf
n}$.  Then $S$ is related to $\omega$ and the torsion $\tau$ by $-S =
\tau + \omega$.  Taking $S = 0$, we obtain the case considered by 
 da Costa in Ref.~\cite{daCosta81}.  We next define an angular
 momentum operator $\Lambda$ in the tangential direction by $\Lambda =
 (u^1 \pi_2 - u^2 \pi_1)/2$. The constrained Hamiltonian is then

\begin{equation}
H_\parallel = K_\parallel + V_{ex},
\label{r79}
\end{equation}
where 

\begin{eqnarray}
K_\parallel 
& = & {1 \over 2} (\pi_\parallel + 2S \langle\Lambda\rangle)^2, 
\label{r110} \\
V_{ex}     
& = & -{\hbar^2 \over 8} {\kappa^2} 
+ 2 S^2( \langle \Lambda^{2} \rangle 
- \langle\Lambda\rangle^2),
\label{r80}
\end{eqnarray}
and where the bracket notation $\langle \;\; \rangle$ denotes the
expectation value with respect to the transverse mode $\chi$.  

Observe that the tangential kinetic energy $K_\parallel$ departs from
$\pi_\parallel^2/2$ due to the inclusion of the term $2 S \langle
\Lambda
\rangle$, which couples the angular momentum of the transverse mode to
the rate of potential twist.  This term is a gauge coupling, a fact we
explore further in Sect.~\ref{s13}.  For now, we simply note that
because the curve is one-dimensional, the gauge coupling can be
removed from Eq.~(\ref{r110}) by a gauge transformation.  In the
present context, a gauge transformation consists of changing the phase
of the wave function $\phi$.  This transformation is not without its
consequences, however, as it will obviously change the boundary
conditions which $\phi$ must satisfy.  Also, we stress that if the
constraint manifold has dimension greater than one, it will not in
general be possible to remove the gauge coupling by a gauge
transformation.  

As a final observation on $K_\parallel$, notice that the gauge
coupling is of order $\hbar^0$, which indicates that it is essentially
a classical quantity.  This coupling should therefore appear in a
classical theory of constraints which takes into account the potential
twist.

Turning to the quantity $V_{ex}$, we note that it is a real-valued
function of $\alpha$, containing no differential operators.  For this
reason, we call $V_{ex}$ an extrapotential.  The extrapotential
contains two terms, $-\hbar^2 \kappa^2/8$ and $ 2 S^2(\langle
\Lambda^2\rangle - \langle \Lambda \rangle^2)$.  The first of these
 was derived by da Costa for the case $S = 0$~\cite{daCosta81}.  It
 has the physical effect of attracting $\phi$ to regions of high
 curvature, a fact which may produce curvature-induced bound states in
 the waveguide.  Such bound states are of current interest
\cite{Exner89a,Exner89b,Ashbaugh90} and are reviewed by Duclos and Exner
\cite{Duclos95}.  The term  $-\hbar^2 \kappa^2/8$ is of order  $\hbar^2$ and therefore disappears in the classical (and semi-classical)
limit.  The second term of $V_{ex}$, like the gauge coupling in
$K_\parallel$, depends on both the potential twist $S$ and the angular
momentum $\Lambda$.  Notice, however, that it is the standard
deviation of the angular momentum which appears in $V_{ex}$.  This
means, for example, that the second term of $V_{ex}$ vanishes for
transverse modes which are angular momentum eigenstates.  It is
interesting to observe that, in contrast to the first term, the second
term of $V_{ex}$ has the physical effect of expelling the wave
function $\phi$ from regions of high twist $S$.  Also, the second term
is of order $\hbar^0$, which means that, like the gauge coupling in
$K_\parallel$, it survives the classical limit.

\section{Introduction to the General Problem}

\label{s14}

We describe here how the setup in Sect.~\ref{s7} is modified for the
general problem.  First, the ambient space in Sect.~\ref{s7} was
assumed to be ${\Bbb R}^3$.  In the general problem, we allow the
ambient space to be an arbitrary Riemannian manifold, which we denote
by ${\cal A}$.  The kinetic energy of the wave function $\psi$,
defined over ${\cal A}$, is given by $K = -\hbar^2 \triangle /2$,
where $\triangle$ is the Laplacian on ${\cal A}$.  Unlike
Sect.~\ref{s7}, the ambient space is not assumed to be flat, and, as
we will discover, the curvature of the ambient space produces
additional terms in $V_{ex}$.

Next, we constrain the wave function to lie in the vicinity of a
(locally) arbitrary (embedded) submanifold ${\cal C}$ of ${\cal A}$
with dimension $m$ and codimension $d$. We call ${\cal C}$ the
constraint manifold.  In Sect.~\ref{s7}, the constraint manifold was a
one-dimensional curve.  The curvature and torsion of this curve played
a critical role in the analysis.  The appropriate generalization of
the curvature and torsion is the second fundamental form $T$, which is
a rank three tensor.  (See Appendix~\ref{s1}.)

In Sect.~\ref{s7}, the constraint was imposed by a hard-wall potential
that was infinite outside of a tube and zero inside.  We then took the
limit in which the width of the tube went to $0$.  In the general
problem, we impose the constraint by an arbitrary potential $V_\perp$,
subject to a few reasonable conditions.  This potential is defined on
a set of coordinates transverse to the constraint manifold and, for
this reason, is called the transverse (or constraining) potential.
The transverse potential depends on a scaling parameter $\epsilon$
which is analogous to the width parameter of the tube; the constraint
is imposed by taking the limit $\epsilon$ goes to $0$.

One of the conditions we do still require of $V_\perp$ is that it be
independent of the location on the constraint manifold.  This
condition, as well as a few other minor conditions, are
explained fully in the next section.

\section{The Transverse Potential}

\label{s15} 
\subsection{Constancy of the Transverse Potential}

\label{s5}

In Sect.~\ref{s7}, we defined the constraining potential by first
specifying the form of the potential on a plane normal to the curve
and then specifying the orientation of this potential at all points
along the curve.  For the general case, we use the same fundamental
idea except that now, due to the curvature of
the ambient space, we must take care to define how we generalize the
concept of the normal plane.

It is useful to consider two separate but related spaces for a given
point $q$ on the constraint manifold.  The first is the linear space
of all vectors normal to the constraint manifold. We call this the
normal space at $q$ and denote it by $N_q$.  The second space of
interest is the submanifold of the ambient space formed by geodesics
emanating from $q$ normal to the constraint manifold.  We call this
the transverse space at $q$ and denote it by ${\cal U}_q$.  The spaces
$N_q$ and ${\cal U}_q$ are related by the exponential map which takes
a vector ${\bf v} \in N_q$ into the point $\exp {\bf v} \in {\cal A}$.
The point $\exp {\bf v} \in {\cal A}$ lies on the geodesic emanating
from $q$ in the direction of ${\bf v}$; it lies at a distance $|{\bf
v}| = (\langle {\bf v}, {\bf v} \rangle )^{1/2}$ from $q$ along this
geodesic.  (We use the notation $\langle \; , \; \rangle$ for the
metric on ${\cal A}$.)  Thus, we find ${\cal U}_q = \exp N_q$.  We now
modify our original definition of ${\cal U}_q$.  If the geodesics
emanating from the constraint manifold ${\cal C}$ in the neighborhood
of $q$ flow to an arbitrary length, they will in general intersect one
another.  This can be witnessed even in the simple example of
Sect.~\ref{s7}.  Thus, in defining ${\cal U}_q$, we assume that the
geodesics flow for a small enough length to avoid such intersections
and that this maximal length is independent of the point $q$ on the
constraint manifold.  In summary, then, we foliate a neighborhood
(which we call the tubular neighborhood) of the constraint manifold
${\cal C}$ by the transverse spaces ${\cal U}_q$, which we have in
turn related to the normal spaces $N_q$ by the exponential map.  Using
the exponential map to construct tubular neighborhoods in this fashion
is a standard technique.  For details, see, for example,
Lang~\cite{Lang99} and Vanhecke~\cite{Vanhecke88}. 

Since we have identified normal vectors with points in the
neighborhood of the constraint manifold, we view the transverse
potential $V_\perp$ as a function defined on the normal spaces.  With
this interpretation, we will require that $V_\perp$, as a function of
$q$ and the vectors in $N_q$, be independent of $q$.  By independent,
we really mean independent modulo $SO(d)$ rotations in $N_q$.  Let us
make this more precise.  As in Sect.~\ref{s7}, we specify the
orientation of the transverse potential by an orthonormal basis ${\bf
E}_\mu$, $\mu = 1,...,d$ of the normal space $N_q$.  This basis forms a
normal frame for the constraint manifold which we call the potential
frame.  For a given normal vector field ${\bf u}$, we introduce the
components $u^\mu$, $\mu = 1,...,d$ with respect to ${\bf E}_\mu$.  The
quantities $u^\mu$ coordinatize both the normal space $N_q$ and the
transverse space ${\cal U}_q$, for which they are commonly called
Riemannian normal coordinates \cite{Misner73}.  We use sans serif for
the list of coordinates ${\sf u} = (u^1, ..., u^d)$, reserving the
bold notation ${\bf u}$ for the vector field.  The neighborhood of
${\cal C}$ is therefore conveniently parameterized by $({\sf
u}, q)$.  The heuristic constraint that $V_\perp$ be independent of
position on the constraint manifold can now be made precise by the
following statement: the transverse potential $V_\perp({\sf u}, q)$ as a
function of $({\sf u}, q)$ is required to be independent of $q$.

In general, the construction of the parameters $u^\mu$ presented here
is only possible locally on ${\cal C}$.  That is, it may be impossible
to define $u^\mu$ in the neighborhood of the whole constraint manifold
simultaneously.  The construction can break down in two ways.  First,
it may be impossible to construct a tubular neighborhood for the
entire constraint manifold.  One can see this even with the simple
example of Sect.~\ref{s7}.  If the one-dimensional curve spirals in on
itself, then the width of the tubular neighborhood is forced to go to
$0$.  (Recall that the width of the tubular neighborhood must be the
same for all point on the constraint manifold.)  Assuming that a
tubular neighborhood does indeed exist for the manifold, there is
still a second way in which the construction may break down.  This
occurs if there does not exist a potential frame ${\bf E}_\mu$ which
is globally defined.  (This happens when the normal bundle is
nontrivial.)  For example, let the ambient space be a M\"{o}bius strip
and let the constraint manifold be a curve which wraps around the
M\"{o}bius strip once.  Clearly, there does not exist a normal frame
for ${\cal C}$ which is defined globally. It is our viewpoint that
these two obstacles (in particular the first) are not common in
physical problems.  Even if one did encounter a problem in which the
$u^\mu$ were not definable globally, since the results of this paper
involve only the local form of the Hamiltonian, they would still apply
to such problems.

\subsection{The Potential Twist Tensor}

In this section, we generalize the potential twist $S$, of
Sect.~\ref{s7}, to a rank three potential twist tensor (also denoted
$S$) defined for any $q \in {\cal C}$.  For an arbitrary vector ${\bf
e} \in T_q{\cal A}$, $S_{\bf e}$ is a linear map on $T_q{\cal A}$.
(Here, $T_q{\cal A}$ is the $(d + m)$-dimensional tangent space of
${\cal A}$ at $q$.)  Let ${\bf x} \in T_q {\cal A}$ be an arbitrary
vector tangent to ${\cal C}$.  Then, we define

\begin{equation}
S_{\bf e} {\bf x} = 0.
\label{r72}
\end{equation}
Now let ${\bf v}\in T_q{\cal A}$ be an arbitrary vector normal to ${\cal C}$.  We extend ${\bf v}$ to a vector field on ${\cal C}$ (defined in the
neighborhood of $q$) by assuming that ${\bf v}$ is normal to ${\cal C}$
and furthermore that its components with respect to ${\bf E}_\mu$ are constant.
We now complete the definition of $S_{\bf e}$ by prescribing

\begin{equation}
S_{\bf e} {\bf v} 
=  P_\perp \nabla_{P_\parallel {\bf e}} {\bf v}, 
\label{r27}
\end{equation}
where $\nabla$ is the Levi-Civita connection
\cite{Kobayashi63,Spivak79b,Eguchi80} on ${\cal A}$ and $P_\perp$ and
$P_\parallel$ are the projection operators onto the normal and tangent
spaces of ${\cal C}$ respectively.  It is straightforward to verify
that $S$ defined by Eqs.~(\ref{r72}) and (\ref{r27}) is indeed a
tensor.

Like the second fundamental form $T$ (see
Appendix~\ref{s1}), $S$ satisfies the antisymmetry property

\begin{equation}
\langle {\bf d}, S_{\bf e} {\bf f} \rangle 
= - \langle {\bf f}, S_{\bf e} {\bf d} \rangle,
\label{r114}
\end{equation}
where ${\bf d},{\bf e},{\bf f} \in T_q{\cal A}$ are arbitrary.  To
prove the above equation, we need only consider the case ${\bf d}={\bf
v} \in N_q$, ${\bf f} = {\bf w}\in N_q$, and ${\bf e} = {\bf x}$
tangent to ${\cal C}$, since all other cases vanish.  Since $S$ is a
tensor, we may assume that ${\bf v}$ and ${\bf w}$ are vector fields
and that their components with respect to ${\bf E}_\mu$ are constant.
Since the frame ${\bf E}_\mu$ is orthonormal, $\langle {\bf v}, {\bf
w}
\rangle$ is constant, and therefore  Eq.~(\ref{r27}) implies $\langle {\bf v}, S_{\bf x} {\bf w} \rangle = 
\langle {\bf v}, \nabla_{\bf x} {\bf w} \rangle = 
- \langle \nabla_{\bf x} {\bf v}, {\bf w} \rangle = - \langle {\bf
w},S_{\bf x} {\bf v} \rangle$.

\subsection{Scaling of the Transverse Potential}

\label{s10}

In Sect.~\ref{s7}, we imposed the constraint by taking the limit in
which the width of the waveguide went to zero.  Here, we describe a
similar scaling procedure using, however, a more general transverse
potential.  Heuristically, we assume that $V_\perp({\sf u};\epsilon)$
depends on the scaling parameter $\epsilon$ in such a way that the
potential grows narrower and deeper as $\epsilon$ tends toward $0$.
To make this precise, we rescale the transverse potential in the
following way

\begin{equation}
\tilde{V}_\perp(\tilde{\sf u}; \epsilon)
= \epsilon^2 V_\perp(\epsilon \tilde{\sf u}; \epsilon),
\label{r54}
\end{equation}
where 

\begin{equation}
u^\mu = \epsilon \tilde{u}^\mu.
\label{r52}
\end{equation}
We assume that $\tilde{u}^\mu$ has no dependence itself on $\epsilon$
and that $\tilde{V}_\perp(\tilde{\sf u}; \epsilon)$ is smooth in
$\epsilon$ at $\epsilon =0$, by which we mean that
$\tilde{V}_\perp(\tilde{\sf u};
\epsilon)$ can be expanded as $\tilde{V}_\perp(\tilde{\sf u};
\epsilon) =
\tilde{V}_{\perp}^0(\tilde{\sf u})
+ \epsilon \tilde{V}_{\perp}^1(\tilde{\sf u}) + \epsilon^2
\tilde{V}_{\perp}^2(\tilde{\sf u}) + ..$.  We also assume  that $\tilde{V}_{\perp}^0(\tilde{\sf u})$ does not vanish.  In Sect.~\ref{s24}, we will
make some very natural, further assumptions on the existence of bound
states for the transverse potential and on the smoothness in
$\epsilon$ of the corresponding eigenenergies.

As a concrete example take $\tilde{V}_\perp(\tilde{\sf u}; \epsilon) =
\tilde{V}_\perp(\tilde{\sf u})$ to be a finite-depth square well with no
$\epsilon$ dependence.  Then $V_\perp({\sf u}; \epsilon)$ is a
finite-depth square well whose width scales as $\epsilon$ and whose
depth scales as $1/\epsilon^2$.  Of course, these scaling rules apply
to any potential satisfying the conditions described above.  They
guarantee that the expectation value of $u^\mu$ with respect to a
transverse mode scales as $\epsilon$ (assuming a fixed quantum number
for the transverse mode).  This fact shows that the wave function
becomes more and more localized in the vicinity of the constraint
manifold as $\epsilon$ tends toward $0$.

\section{Expansion of the Kinetic Energy}

\label{s11}

The derivation of the constrained Hamiltonian (such as Eqs.~(\ref{r79})
-- (\ref{r80})) proceeds in two steps.  The first is to expand the
kinetic energy in powers of $\epsilon$.  The second is to transform to a basis
of transverse modes and to apply a first order perturbation treatment
to the expanded Hamiltonian.  This section is devoted to the first
step.

\subsection{Definition of the Vielbein}

\label{s6}

We will express the kinetic energy in terms of a vielbein ${\bf E}_a$,
$a = 1, ..., d + m$, on ${\cal A}$.  Appendix~\ref{s9} gives the
necessary background for this technique.  To span the transverse
dimensions, we take ${\bf E}_\mu = \partial /\partial u^\mu$, $\mu =
1, ..., d$, where it is understood that, for the purpose of the partial
derivative, the position $q \in {\cal C}$ is held fixed.  In selecting
vector fields to span the remaining dimensions, we first choose an
arbitrary set of orthonormal vector fields ${\bf E}_i$, $i = d +
1,...,d + m$, defined over ${\cal C}$ and tangent to ${\cal C}$.  We
then use ${\bf E}_\mu$ to Lie transport these vector fields into the
tubular neighborhood of ${\cal C}$.  That is, we require the Lie
derivatives with respect to ${\bf E}_\mu$ to vanish,

\begin{equation}
[{\bf E}_\mu, {\bf E}_i ] = 0.
\label{r6}
\end{equation}

We use the following notational scheme in this paper.  The indices
$a,b,c,...$ range from $1, ..., d + m$ and label the basis vectors
${\bf E}_a$ and any components with respect to this basis.  The
indices $\mu, \nu, \sigma, ...$ range from $1, ..., d$ and label the
vector fields ${\bf E}_\mu = \partial / \partial u^\mu$ and their
related components.  The indices $i, j, k, ...$ range from $d + 1,
..., d + m$ and label the vector fields ${\bf E}_i$ and their related
components.  Except where otherwise noted, we employ the convention
that an index $a,b, c,...$, $\mu, \nu, \sigma, ...$, or $i,j, k,...$
is implicitly summed over when occurring twice in the same expression.

For future reference, we present some facts regarding the structure
constants $\beta^c_{ab}$, defined by $[ {\bf E}_a, {\bf E}_b ] = \beta_{ab}^c
{\bf E}_c$.   First,
Eq.~(\ref{r6}) immediately yields

\begin{equation}
\beta^c_{\mu j} = \beta^c_{j \mu} = 0.
\label{r1}
\end{equation}
Furthermore, since ${\bf E}_\mu = \partial / \partial u^\mu$ is a
coordinate basis on the transverse spaces ${\cal U}_q$, we find
$\beta^c_{\mu \nu } = 0$.  Combining this with Eq.~(\ref{r1}), we have

\begin{equation}
\beta^c_{\mu b } = \beta^c_{b \mu} = 0.
\label{r81}
\end{equation}

Next we note that
\begin{equation}
0 
= [ {\bf E}_\mu , [ {\bf E}_i, {\bf E}_j ] ] 
= [ {\bf E}_\mu , \beta_{ij}^c {\bf E}_c ] 
= [ \partial / \partial u^\mu  \beta_{ij}^c ] {\bf E}_c, 
\end{equation}
where the first equality follows from the Jacobi identity and
Eq.~(\ref{r6}) and the third equality follows from Eq.~(\ref{r81}).
We use the bracket notation $[ \; \; ]$ in the final equality to
emphasize that the differential operator acts only on the quantities
inside the bracket.  Since the ${\bf E}_c$ form a basis, we have

\begin{equation}
{\partial \beta_{ij}^c \over \partial u^\mu} = 0.
\label{r82}
\end{equation}  
Furthermore, since the ${\bf E}_i$ are tangent to ${\cal C}$ when
restricted to ${\cal C}$, we have $\beta^\sigma_{ij} = 0$ on ${\cal
C}$.  Combining this fact with Eq.~(\ref{r82}), we find that

\begin{equation}
\beta^\sigma_{ij} = 0
\end{equation}
within the tubular neighborhood.  Thus, the distribution of vector
fields ${\bf E}_i$ is integrable everywhere.  (The submanifolds thus
defined by Frobenius's Theorem are manifolds of constant $V_\perp$.)

\subsection{Transformation of the Kinetic Energy}

As in Appendix~\ref{s9}, the momentum operators are defined to be $\pi_a
= -i\hbar{\bf E}_a$.  They are not in general Hermitian since the
Hermitian conjugate is given by Eq.~(\ref{r74}).  The kinetic energy
is given by $K = \pi_a^\dagger G^{ab} \pi_b/2$, where $G_{ab} =
\langle {\bf E}_a, {\bf E}_b \rangle$ are the components of the metric
tensor and $G^{ab}$ is the inverse of $G_{ab}$.

Appendix~\ref{s9} also provides the framework for scaling the quantum
wave function by an arbitrary (strictly) positive function $s : {\cal A} \rightarrow {\Bbb R}$ (see
Eq.~(\ref{r83})) in order to modify the form of the kinetic energy.
We apply this scaling formalism here, taking

\begin{equation}
s = G^{1/4},
\end{equation}
where $G = \det G_{ab}$.  As mentioned in Appendix~\ref{s9},
this scaling  defines a new inner product of wave functions.  We observe that the
original inner product of two wave functions $\varphi$ and $\varphi'$
is given by

\begin{equation}
\langle \varphi | \varphi' \rangle 
= \int \sqrt{G} \nu \; \varphi^*({\sf u}, q) \varphi'({\sf u}, q),
\end{equation}
where $\nu$ is the $(d + m)$-form

\begin{equation}
\nu 
= E^{*1} \wedge ... \wedge E^{*(d + m)}
= du^1 \wedge ... \wedge du^d \wedge E^{*(d + 1)} 
\wedge ... \wedge E^{*(d + m)}.
\label{r141}
\end{equation}
Here, $E^{* a}$ is the basis of one-forms dual to the vielbein ${\bf
E}_a$.  We have also used the fact that $E^{* \mu} = d u^\mu$.  (Be
careful not to confuse the notation $\langle \; | \; \rangle$ with
$\langle \; , \; \rangle$ which denotes the Riemannian metric on
${\cal A}$.)  From Eq.~(\ref{r140}) we therefore find that the scaled
inner product of two (scaled) wave functions $\psi$ and $\psi'$ is

\begin{equation}
\langle \psi | \psi' \rangle_s 
= \int \nu \; \psi^*({\sf u}, q) \psi'({\sf u}, q).
\end{equation}
This scaled inner product gives rise to a scaled Hermitian conjugate,
denoted $\dagger(s)$.  Using Eqs.~(\ref{r74}), (\ref{r78}), and
(\ref{r81}), we find that $\pi_\mu$ is Hermitian with respect to the
scaled Hermitian conjugate,

\begin{equation}
\pi_\mu^{\dagger(s)} = \pi_\mu.
\label{r20}
\end{equation}
Furthermore, Eqs.~(\ref{r74}), (\ref{r78}), and (\ref{r1}) 
give the (scaled) Hermitian conjugate of $\pi_j$ as

\begin{equation}
\pi_j^{\dagger(s)} 
= \pi_j + i \hbar \beta^b_{j b} = \pi_j + i \hbar \beta^k_{jk}.
\label{r2}
\end{equation}

We now restrict the momentum operator $\pi_j$ to ${\cal C}$.  We write
$\pi_j|_0$ to make this explicit; we use the notation $|_0$ for any
quantity restricted to ${\cal C}$ since this corresponds to $u^\mu =
0$.  For present purposes, we consider the constraint manifold in its
own right without being viewed as embedded in the ambient space.  With
this interpretation, the vector field $\pi_j|_0$ has a well-defined
Hermitian conjugate which we denote by $(\pi_j|_0)^\dagger$.  This
Hermitian conjugate is given by Eq.~(\ref{r74}), where it is
understood that the symbol $G$ now refers only to the determinant of
the metric $G_{ij}$ on ${\cal C}$.  However, since the basis ${\bf E}_i|_0$ is
orthonormal, we have $G = 1$, and hence

\begin{equation}
(\pi_j|_0)^{\dagger} 
= \pi_j|_0  + i \hbar \gamma^k_{jk},
\label{r90}
\end{equation}
where the functions $\gamma^k_{ij}$, defined on ${\cal C}$, are the
structure constants for ${\bf E}_i|_0$.  Since $[ {\bf E}_i|_0, {\bf
E}_j|_0] = [ {\bf E}_i, {\bf E}_j]|_0$, the structure constants
$\gamma^k_{ij}$ are equal to $\beta^k_{ij}|_0$.  Comparing
Eq.~(\ref{r90}) to Eq.~(\ref{r2}), we now have the following convenient
description for $\pi_j^{\dagger(s)}$ when restricted to ${\cal C}$

\begin{equation}
\left. \left(\pi_j^{\dagger(s)}\right)\right|_0 
= (\pi_j|_0)^\dagger.
\label{r91}
\end{equation}

The scaled kinetic energy is given by Eq.~(\ref{r76}).  Noting
Eq.~(\ref{r20}), we rewrite this as

\begin{equation}
K_s 
= {1\over 2}\left( \pi_\mu G^{\mu \nu}\pi_\nu + \pi_\mu G^{\mu j} \pi_j 
+ \pi_i^{\dagger(s)} G^{i\nu} \pi_\nu + \pi_i^{\dagger(s)} G^{ij} \pi_j \right) + V_s
\label{r3}
\end{equation}
where, 

\begin{equation}
V_s 
=  -\frac{1}{8} 
\left( {1\over 4}G^{ab} [ \pi_a \ln G] [\pi_b \ln G]
+ [ \pi_a^{\dagger (s)} G^{ab} [ \pi_b \ln G ] ]\right). 
\label{r15}
\end{equation}
We will henceforth drop the $s$ index on $K_s$, $\dagger(s)$, and $\langle \; \; | \; \; \rangle_s$, with the scaling being implicitly understood.

\subsection{Expansion of the Kinetic Energy}

\label{s19}

In this section, we expand the kinetic energy Eq.~(\ref{r3}) through
order $\epsilon^0$.  Recall from Eq.~(\ref{r52}) that $u^\mu$ is of
order $\epsilon^1$, and hence the momentum $\pi_\mu = -i
\hbar
\partial / \partial u^\mu$ is of order $\epsilon^{-1}$. From Eq.~(\ref{r6}), we see that the momentum $\pi_i$ is of order
$\epsilon^0$.  Furthermore, from Eq.~(\ref{r82}), we see that
$\beta^k_{ij}$ is of order $\epsilon^0$, and combining this fact with
Eq.~(\ref{r2}) we find that $\pi_i^{\dagger}$ is of order
$\epsilon^0$.  (Recall that the index ``$(s)$'' is now implicit.)
These scaling properties imply that to expand Eq.~(\ref{r3}) to order
$\epsilon^0$, we must expand $V_s$, $G^{ij}$, $G^{i\mu}$, and $G^{\mu
\nu}$ to orders $\epsilon^0$, $\epsilon^0$, $\epsilon^1$, and $\epsilon^2$
respectively.

Since ${\bf E}_a$ is an orthonormal frame at $u^\mu = 0$, we have
the following identities

\begin{eqnarray}
G_{ab}|_0 
& = & G^{ab}|_0 
= \delta_{ab}, 
\label{r84}\\
G^{ab}_{\; \; \; \; ,\sigma}|_0
& = & - G_{ab,\sigma}|_0, 
\label{r85}
\end{eqnarray}
where we use the notation ``$,\sigma$'' for the derivative $\partial /
\partial u^\sigma$.  Equations~(\ref{r84}) and (\ref{r85}) yield the
following expansions of $G^{ab}$,

\begin{eqnarray}
G^{ij}({\sf u}) & = & \delta_{ij} + O(\epsilon), 
\label{r86} \\ 
G^{\mu j}({\sf u}) 
& = & G^{\mu j}_{\; \; \; \; , \sigma}|_0 u^\sigma + O(\epsilon^2) 
= - G_{\mu j, \sigma}|_0 u^\sigma +O(\epsilon^2), \\
G^{\mu \nu}({\sf u})
& = & \delta_{\mu \nu} + G^{\mu \nu}_{\; \; \; \; ,\sigma}|_0 u^\sigma 
+ {1\over 2} G^{\mu \nu}_{\; \; \; \; ,\sigma \tau}|_0 u^\sigma u^\tau 
+ O(\epsilon^3) \nonumber \\
& = & \delta_{\mu \nu} - G_{\mu \nu ,\sigma}|_0 u^\sigma 
+ {1\over 2} (- G_{\mu \nu ,\sigma \tau} 
+ 2 G_{\mu a,\sigma} G_{a \nu, \tau})|_0 u^\sigma u^\tau + O(\epsilon^3).
\label{r87}
\end{eqnarray}

The derivatives of the metric $G_{ab}$ appearing above are
conveniently expressed in terms of the potential twist tensor $S$ and
the Riemannian curvature $R$ on ${\cal A}$.  To see this, we first
introduce the components of $S$, $T$, and $R$ via

\begin{eqnarray}
S_{abc} & = & \langle {\bf E}_a, S_{{\bf E}_c} {\bf E}_b \rangle, 
\label{r93} \\
T_{abc} & = & \langle {\bf E}_a, T_{{\bf E}_c} {\bf E}_b \rangle, 
\label{r95} \\
R_{abcd} & = & \langle {\bf E}_a, R_{{\bf E}_c {\bf E}_d} {\bf E}_b \rangle,
\end{eqnarray}
where $T$ is included for completeness and for future reference.  We
then have the following identities

\begin{eqnarray}
G_{\mu j, \sigma}|_0 & = & S_{\mu \sigma j}|_0, 
\label{r7} \\
G_{\mu \nu, \sigma}|_0 & = & 0, 
\label{r9}\\
G_{\mu \nu, \sigma \tau}|_0 
&= & -{1 \over 3} 
\left( R_{\mu \sigma \nu \tau} + R_{\nu \sigma \mu \tau} \right)|_0.
\label{r10}
\end{eqnarray}
In actuality, the $|_0$ notation on $S_{\mu \sigma j}$ is redundant
since $S$ is only defined on ${\cal C}$, but we will make use of this
notation as a convenient reminder.  We will derive Eqs.~(\ref{r7}) --
(\ref{r10}) momentarily, but for now we insert them into
Eqs.~(\ref{r86}) -- (\ref{r87}) to obtain

\begin{eqnarray}
G^{ij}({\sf u}) & = & \delta_{ij} + O(\epsilon), 
\label{r12} \\ 
G^{\mu j}({\sf u}) 
& = &  - S_{\mu \sigma j}|_0 u^\sigma +O(\epsilon^2), 
\label{r13} \\
G^{\mu \nu}({\sf u})
& = & \delta_{\mu \nu} 
+ \left.\left( {1\over 3} R_{\mu \sigma \nu \tau}  
+ S_{\mu \sigma k} S_{\nu \tau k} \right)\right|_0 u^\sigma u^\tau + O(\epsilon^3),
\label{r14}
\end{eqnarray}
where we have used the well-known symmetry of the Riemannian
curvature $R_{abcd} = R_{cdab}$.  We next insert Eqs.~(\ref{r12}) --
(\ref{r14}) into Eq.~(\ref{r3}) to arrive at the main result of this
section,

\begin{equation}
K =  K_\perp + K_\parallel^p + K_R + V_{ex}^p + O(\epsilon),
\label{r32}
\end{equation}
where

\begin{eqnarray}
K_\perp & = & {1\over 2} \pi^\mu|_0 \pi_\mu, 
\label{r88} \\
K_\parallel^p 
& = & {1\over 2} (\pi^i|_0 +  S^{\mu \nu i}|_0 \Lambda_{\mu \nu})^{\dagger} (\pi_i +  S^{\sigma \tau}_{\;\; \; \; i}|_0 \Lambda_{\sigma \tau}), 
\label{r55} \\
K_R & = & {1\over 6}  R^{\mu \nu \sigma \tau}|_0 \Lambda_{\mu \nu} \Lambda_{\sigma \tau}, 
\label{r89} \\
V_{ex}^p & = & V_s |_0.
\end{eqnarray}
We have taken advantage of the antisymmetry property Eq.~(\ref{r114}),
$S_{\mu \nu i} = - S_{\nu \mu i}$, and the well-known antisymmetry
relations $R_{\mu \nu \sigma \tau} = -R_{\nu \mu \sigma
\tau} = - R_{\mu \nu \tau \sigma}$ to introduce the operators

\begin{equation}
\Lambda_{\mu \nu} 
= {1 \over 2} (u_\mu \pi_\nu - u_\nu \pi_\mu)
= {1 \over 2} (\pi_\nu u_\mu - \pi_\mu u_\nu),
\label{r46}
\end{equation}
which are generalized angular momentum operators acting on the
transverse space.  That is, they generate $SO(d)$ rotations in the
transverse space.  They are the generalization of the
angular momentum $\Lambda$ defined in Sect.~\ref{s7}.  

In Eqs.~(\ref{r88}) -- (\ref{r89}), we have employed the standard
practice of raising tensor indices by contraction with $G^{ab}$.
Thus, $\pi^\mu = G^{\mu a}\pi_a$, $S^{\mu \nu i} = G^{\mu a} G^{\nu b}
G^{i c} S_{abc}$, etc.  However, since $G^{ab}|_0 = \delta_{ab}$,
the raised components and lowered components of any tensor evaluated
at $u^\mu = 0$ are actually equal.  One could, therefore, equally well
have written Eqs.~(\ref{r88}) -- (\ref{r89}) with all components
lowered.  The purpose of using raised components is simply to express
these equations in manifestly covariant form.

We now mention a few facts concerning the Hermitian conjugate which we used
to derive Eq.~(\ref{r55}).  First, notice from Eq.~(\ref{r20}) that
$\Lambda_{\mu \nu}^{\dagger} = \Lambda_{\mu
\nu}$.  Also notice that since $S^{\mu \nu i}|_0$ has no dependence on
$u^\mu$, $S^{\mu \nu i}|_0$ and $\Lambda_{\mu \nu}$ commute.  This
means in particular that the Hermitian conjugate in Eq.~(\ref{r55})
may be applied to the $\pi^i|_0$ term alone.  Finally, we used
Eq.~(\ref{r91}) to relate $\left(\pi_i|_0 \right)^\dagger$ to
$( \pi_i^\dagger )|_0$.  Notice from
Eq.~(\ref{r90}) that if ${\bf E}_i|_0$ is a coordinate basis on ${\cal
C}$ then the Hermitian conjugate may be dispensed with altogether.

We call the terms $K_\perp$, $K_\parallel^p$ and $K_R$ appearing in
Eq.~(\ref{r32}) the transverse kinetic energy, the (preliminary)
tangential kinetic energy, and the curvature energy respectively.  The
last term $V_{ex}^p$, being a scalar, nondifferential operator, we call
the (preliminary) extrapotential.  The three terms $K_\parallel^p$, $K_R$, and
$V_{ex}^p$ are all order $\epsilon^0$.  The transverse kinetic energy
$K_\perp$ is of order $\epsilon^{-2}$ and therefore blows up as
$\epsilon$ shrinks to $0$.  The energy associated with this term will
therefore be subtracted off with the remaining three terms giving rise
to the residual kinetic energy.  Notice that each of the four terms in
Eq.~(\ref{r32}) is Hermitian with respect to the (scaled) Hermitian
conjugate.

\subsection{Proof of Identities~(\ref{r7}) -- (\ref{r10})}

\label{s20}

We return now to justify Eqs.~(\ref{r7}) -- (\ref{r10}).  Considering 
Eq.~(\ref{r7}), we have

\begin{eqnarray}
G_{\mu j,\sigma}|_0 
& = & \left( \nabla_{{\bf E}_\sigma} G_{\mu j}\right)|_0 
= \left. \left( \nabla_{{\bf E}_\sigma} 
\left \langle {{\bf E}_\mu}, 
{\bf E}_j \right \rangle\right)\right|_0 \nonumber \\
& = & \left. \left \langle \nabla_{{\bf E}_\sigma}  
{{\bf E}_\mu}, 
{\bf E}_j \right \rangle \right|_0 
+ \left. \left \langle {{\bf E}_\mu}, 
\nabla_{{\bf E}_\sigma}  {\bf E}_j \right \rangle \right|_0,
\label{r4}
\end{eqnarray}
where in the first equality, we replaced the coordinate derivative by
the covariant derivative, treating $G_{\mu j}$ as a scalar function.
The second equality is the definition of $G_{\mu j}$, and the third equality
follows from the Leibniz rule and the vanishing of the metric tensor under
covariant differentiation.  We next define the vector

\begin{equation}
{\bf M}_{\sigma \mu} 
= \nabla_{{\bf E}_\sigma}  
{\bf E}_\mu, 
\label{r92}
\end{equation}
which we now demonstrate vanishes on ${\cal C}$.  To prove this, we
first show that it is everywhere symmetric in $\mu$ and $\sigma$ by
using the 
general formula

\begin{equation}
\nabla_{\bf d} {\bf e} - \nabla_{\bf e} {\bf d} = [{\bf d}, {\bf e}],
\label{r8}
\end{equation}
where ${\bf d}$ and ${\bf e}$ are arbitrary vector fields on ${\cal
A}$.  By substituting ${\bf d} = {\bf E}_\sigma$ and ${\bf e} = {\bf
E}_\mu$ and recalling that $[ {\bf E}_\mu , {\bf E}_\sigma] = 0$, we
find that ${\bf M}_{\sigma \mu} = {\bf M}_{\mu \sigma}$.  Since ${\bf
M}_{\sigma
\mu}$ is symmetric, it vanishes if and only if $v^\sigma v^\mu {\bf M}_{\sigma \mu}  =
0$ for an arbitrary list of (constant) real numbers $v^\sigma$.
For such an arbitrary list, we define the vector field ${\bf
v} = v^\sigma {\bf E}_\sigma$ over ${\cal A}$.  Since $\partial v^\mu
/ \partial u^\sigma = 0$, we see from Eq.~(\ref{r92}) that $v^\sigma
v^\mu {\bf M}_{\sigma
\mu}  = \nabla_{\bf v} {\bf v}$.  Since the quantities $u^\mu$ are defined via
geodesic flow away from ${\cal C}$, an integral curve of ${\bf v}$
which passes through ${\cal C}$ is itself a geodesic.  By the geodesic
equation, $(\nabla_{\bf v} {\bf v})|_0 = 0$.  Thus, $v^\sigma v^\mu {\bf
M}_{\sigma
\mu}|_0 = 0$ and hence

\begin{equation}
\left. \left( \nabla_{{\bf E}_\sigma}  
{{\bf E}_\mu}  
\right)\right|_0
= 0. 
\label{r28}
\end{equation}
We return to Eq.~(\ref{r4}) and write

\begin{equation}
G_{\mu j,\sigma}|_0 
= \left. \left \langle {{\bf E}_\mu}, 
\nabla_{{\bf E}_\sigma} {\bf E}_j \right \rangle \right|_0
= \left. \left \langle {{\bf E}_\mu}, 
\nabla_{{\bf E}_j} {{\bf E}_\sigma}  \right \rangle \right|_0= S_{\mu \sigma j}|_0,
\end{equation}
where the first equality follows from Eq.~(\ref{r28}), the second from
Eqs.~(\ref{r8}) and (\ref{r6}), and the third from Eqs.~(\ref{r93}) and
(\ref{r27}).

To prove Eqs.~(\ref{r9}) and (\ref{r10}), we fix a point $q$ on the
constraint manifold and restrict our attention to a single transverse
space ${\cal U}_q$ which we temporarily forget is embedded in ${\cal
A}$.  Recall that the vectors ${\bf E}_\mu$ are tangent to ${\cal
U}_q$ and $G_{\mu \nu} = \langle {\bf E}_\mu, {\bf E}_\nu \rangle$ is
the metric tensor on ${\cal U}_q$.  Furthermore, the coordinates
$u^\mu$ are Riemannian normal coordinates on ${\cal U}_q$ and it is
well-known that the expansion of the metric to second order in the
Riemannian normal coordinates is \cite{Misner73}

\begin{equation}
G_{\mu \nu}({\sf u}) 
= \delta_{\mu \nu} - 
{1\over 3}\bar{R}_{\mu \sigma \nu \tau }|_0 u^{\sigma}u^{\tau} + ..., 
\label{r94}
\end{equation}
where $\bar{R}$ is the Riemannian curvature of the transverse space
${\cal U}_q$.  The vanishing in Eq.~(\ref{r94}) of the term linear in
${\sf u}$ proves Eq.~(\ref{r9}).  Similarly, the quadratic term in
Eq.~(\ref{r94}) yields

\begin{equation}
G_{\mu \nu,\sigma \tau}|_0 
= -{1 \over 3} 
\left( \bar{R}_{\mu \sigma \nu \tau} 
+ \bar{R}_{\nu \sigma \mu \tau} \right)|_0.
\end{equation}

To complete the proof of Eq.~(\ref{r10}), we must prove that the
components $\bar{R}_{\mu \sigma \nu \tau}|_0$ of the Riemannian curvature
on ${\cal U}_q$ agree with the components $R_{\mu \sigma \nu
\tau}|_0$ of the Riemannian curvature on ${\cal A}$. To prove this, we
use the Gauss relation given by Eq.~(\ref{r11}) and which we reexpress
here in component form

\begin{equation}
R_{\mu \sigma \nu \tau}
= \bar{R}_{\mu \sigma \nu \tau}
+ \bar{T}^a_{\; \; \sigma \nu} \bar{T}_{a \mu \tau}
- \bar{T}^a_{\; \; \sigma \tau} \bar{T}_{a \mu \nu}.
\label{r96}
\end{equation}
Since we are applying the Gauss equation to the submanifold ${\cal
U}_q$ instead of ${\cal C}$, we place an overbar on the symbols for
the second fundamental form and the Riemannian curvature.  Here,
$\bar{T}$ is the second fundamental form of ${\cal U}_q$.  Recall that
$P_\parallel$ and $P_\perp$ were defined to be respectively the tangent
and normal projection operators onto ${\cal C}$.  We extend the
definition of these operators for $u^\mu$ not equal to
$0$ by defining $P_\parallel$ and $P_\perp$ to be the normal and
tangent projection operators respectively onto ${\cal U}_q$.  With
this definition, the second fundamental form $\bar{T}$ is given by
(see Eq.~(\ref{r25}))

\begin{equation}
\bar{T}_{\bf e} {\bf f} 
= P_\parallel \nabla_{P_\perp {\bf e}} P_\perp {\bf f} 
+  P_\perp \nabla_{P_\perp {\bf e}} P_\parallel {\bf f}, 
\end{equation}
where ${\bf e}$ and ${\bf f}$ are arbitrary vector fields over ${\cal
U}_q$ which are tangent to ${\cal A}$.  Since ${\bf E}_\mu$ is tangent
to ${\cal U}_q$ everywhere, we have

\begin{equation}
\bar{T}_{\sigma \mu \nu} 
= \left \langle {\bf E}_\sigma, 
\bar{T}_{{\bf E}_{\nu}} {\bf E}_{\mu} \right \rangle
= 0.
\label{r98}
\end{equation}
Furthermore, since ${\bf E}_i$ is normal to ${\cal U}_q$ at $u^\mu =
0$, we have

\begin{equation}
\bar{T}_{i \mu \nu}|_0 
= \left. \left \langle {\bf E}_i, 
\bar{T}_{{\bf E}_\nu} 
{\bf E}_\mu \right \rangle \right|_0
= \left. \left \langle {\bf E}_i, 
\nabla_{{\bf E}_\nu} {{\bf E}_\mu} \right \rangle \right|_0 
= 0,
\label{r97}
\end{equation}
where the last equality follows from Eq.~(\ref{r28}).  Combining
Eqs.~(\ref{r98}) and (\ref{r97}) yields $\bar{T}^a_{\; \; \mu \nu}|_0
= \bar{T}_{a \mu \nu}|_0 = 0$ from which follows, using
Eq.~(\ref{r96}), $\bar{R}_{\mu \sigma \nu \tau}|_0 = R_{\mu
\sigma \nu \tau}|_0$.  This concludes the proof of Eq.~(\ref{r10}).

\subsection{Computation of the Extrapotential}

\label{s16}
In this section, we analyze the extrapotential $V_{ex}^p$ formed by
evaluating Eq.~(\ref{r15}) at $u^\mu = 0$.  As we will see, $V_{ex}^p$
may be expressed solely in terms of the second fundamental form $T$ of
the constraint manifold and the Riemannian curvature $R$ of ${\cal A}$
evaluated on ${\cal C}$ with no dependence on the potential twist $S$.
Specifically, we will derive the following manifestly covariant form

\begin{equation}
V_{ex}^p = -{\hbar^2\over 8} \left. \left( 
 2 T^{i\mu j}T_{i \mu j} 
- T^{i \mu}_{\; \; \; \; i}T^j_{\; \; \mu j} 
+  2 R^{i \mu}_{\; \; \; \; i \mu}
+{2 \over 3} R^{\mu \nu}_{\; \; \; \; \mu \nu}\right)\right|_0.
\label{r99}
\end{equation}
Setting $R_{abcd} = 0$, the above equation agrees with da Costa
(Ref.~\cite{daCosta82}, Eq.~(33)).  Da Costa also assumes that $S =
0$.  Since we do not make this assumption, Eq.~(\ref{r99}) is a
generalization of da Costa's result to both the case of nonzero
Riemannian curvature in the ambient space and nonzero twist of the
potential.

There are several other convenient forms for $V_{ex}^p$.  We
first introduce the following notation

\begin{eqnarray}
{\cal R} & = & R^{ab}_{\; \; \; \; ab}|_0, \\
{\cal R}_\perp & = & R^{\mu \nu}_{\; \; \; \; \mu \nu}|_0, \\
{\cal R}_\parallel & = & R^{ij}_{\; \; \; \; ij}|_0, \\
\hat{\cal R} & = & \hat{R}^{ij}_{\; \; \; \; ij}, \\
{\cal T}^2 
& = & {1\over 2}T^{abc}T_{abc} 
= T^{i \mu j}T_{i \mu j}
= T^{\mu ij}T_{\mu ij}, 
\label{r21} \\
{\cal M}^2 
& = & T^{ab}_{\; \; \; \; a} T^c_{\; \; bc} 
= T^{i \mu}_{\; \; \; \; i}T^j_{\; \; \mu j}
= T^{\mu i}_{\; \; \; \; i} T_{\mu \; \; j}^{\; \; j},
\label{r22}
\end{eqnarray}
where we use Eqs.~(\ref{r136}), (\ref{r138}), and (\ref{r139}) in
Eqs.~(\ref{r21}) and (\ref{r22}).  The quantities ${\cal R}$ and
$\hat{\cal R}$ are the scalar curvatures on ${\cal A}$ and ${\cal C}$
respectively.  The quantity ${\cal M}$ is called the mean curvature.
Using the fact that ${\cal R} = {\cal R}_\parallel + {\cal R}_\perp + 2 R^{i\mu}_{\; \; \;
\; i\mu}|_0$, we rewrite Eq.~(\ref{r99}) as

\begin{equation}
V_{ex}^p 
= -{\hbar^2\over 8} \left( 
 2 {\cal T}^2 - {\cal M}^2 + {\cal R} - {\cal R}_\parallel 
-{1 \over 3} {\cal R}_\perp \right).
\label{r143}
\end{equation}
Furthermore, the Gauss Eq.~(\ref{r11}) yields

\begin{equation}
{\cal T}^2 = {\cal M}^2 - \hat{\cal R} + {\cal R}_\parallel,
\end{equation}
from which we find

\begin{eqnarray}
V_{ex}^p 
& = & -{\hbar^2 \over 8}
\left( {\cal T}^2 - \hat{\cal R} + {\cal R} - {1\over 3} {\cal R}_\perp \right)
\label{r107} \\
& = & -{\hbar^2 \over 8}
\left( {\cal M}^2 - 2 \hat{\cal R} + {\cal R} + {\cal R}_\parallel 
- {1\over 3} {\cal R}_\perp \right).
\label{r135}
\end{eqnarray}
Assuming the tensor $R = 0$, Eq.~(\ref{r135}) agrees with Ref.~\cite{daCosta82}, Eq.~(36).

The remainder of this section is devoted to the derivation of
Eq.~(\ref{r99}).  Considering the first term of Eq.~(\ref{r15}), we
observe that since $G_{ab}|_0 = \delta_{ab}$, we find

\begin{eqnarray}
\; [ \pi_i \ln G ]|_0 
& = & 0,
\label{r17} \\
\; [\pi_\mu \ln G]|_0 
& = & -i \hbar (G^{ab}G_{ba,\mu})|_0 
= -i \hbar G_{aa,\mu}|_0 
= -i \hbar G_{jj, \mu}|_0,
\label{r100}
\end{eqnarray}
where we have used Eq.~(\ref{r9}) in the last step of
Eq.~(\ref{r100}).  Equations~(\ref{r17}) and (\ref{r100}) yield

\begin{equation}
{1\over 4}\left( G^{ab} [ \pi_a \ln G] [\pi_b \ln G] \right)|_0
= - {\hbar^2 \over 4} \left( G_{ii,\mu} G_{jj,\mu} \right)|_0.
\end{equation}
Considering the second term of Eq.~(\ref{r15}), we note

\begin{equation}
[\pi_a^{\dagger} G^{ab} [ \pi_b \ln G ] ]|_0
=  \left( [\pi_a^{\dagger}, G^{ab}] [ \pi_b \ln G ]
+ G^{ab} [\pi_a^{\dagger} [ \pi_b \ln G ] ]\right)|_0.
\label{r101}
\end{equation}
The first term of Eq.~(\ref{r101}) vanishes from Eqs.~(\ref{r20}),
(\ref{r2}), (\ref{r13}), (\ref{r14}), and (\ref{r17}).  The second
term evaluates to

\begin{eqnarray}
\left(G^{ab} [\pi_a^{\dagger} [ \pi_b \ln G ] ]\right)|_0
& = & [\pi_\mu [ \pi_\mu \ln G ] ]|_0
= \hbar^2 \left( G_{ab,\mu}G_{ab,\mu} - G_{aa,\mu\mu} \right)|_0 \nonumber \\
& = & \hbar^2 \left. \left( 
2S_{\mu \nu i} S_{\mu \nu i} + G_{ij,\mu}G_{ij,\mu} 
+ {2\over 3} R_{\mu \nu \mu \nu} -G_{ii,\mu \mu}\right) \right|_0,
\end{eqnarray}
where the first equality follows from Eqs.~(\ref{r20}), (\ref{r2}),
(\ref{r84}), and (\ref{r17}), the second equality is a
straightforward computation, and the third equality results from
Eqs.~(\ref{r7}) -- (\ref{r10}).  Collecting the preceding results, we
find

\begin{equation}
V_{ex}^p = \left. -{\hbar^2\over 8} \left( -{1\over 4} G_{ii,\mu} G_{jj,\mu} +
G_{ij,\mu} G_{ij,\mu} - G_{ii,\mu \mu} 
+ 2S_{\mu \nu i} S_{\mu \nu i} 
+ {2 \over 3} R_{\mu \nu \mu \nu} \right) \right|_0.
\label{r16}
\end{equation}
The various derivatives of $G_{ij}$ appearing in the above may be
reexpressed using the following identities, to be derived momentarily,
 
\begin{eqnarray}
G_{ij,\mu}|_0 & = & 2T_{i \mu j}|_0, 
\label{r18} \\
G_{ij,\mu \nu}|_0 
& = & \left. (T_{a \mu i} T_{a \nu j} 
+ T_{a \mu j} T_{a \nu i} 
+ S_{a \mu i} S_{a \nu j} 
+ S_{a \mu j} S_{a \nu i} 
- R_{i \mu j \nu} 
- R_{j \mu i \nu} 
) \right|_0.
\label{r19}
\end{eqnarray}  
Upon inserting Eqs.~(\ref{r18}) and (\ref{r19}) into Eq.~(\ref{r16})
one obtains Eq.~(\ref{r99}).

Considering  Eq.~(\ref{r18}), it follows from 

\begin{equation}
G_{ij,\mu}|_0 
= (\nabla_{{\bf E}_\mu} \langle {\bf E}_i, {\bf E}_j \rangle)|_0
= \left. \left \langle \nabla_{{\bf E}_i} {{\bf E}_\mu}, 
{\bf E}_j \right \rangle\right|_0 
+ \left. \left \langle {\bf E}_i, 
\nabla_{{\bf E}_j} {{\bf E}_\mu} \right \rangle\right|_0 
= 2T_{i\mu j}|_0,
\end{equation}
where in the second equality we used the Leibniz rule and interchanged
the derivatives by virtue of Eqs.~(\ref{r8}) and (\ref{r6}).  The
final equality follows from the definition of the second fundamental
form Eq.~(\ref{r25}) and Eqs.~(\ref{r23}) and (\ref{r24}).

Considering Eq.~(\ref{r19}), we have

\begin{eqnarray}
G_{ij,\mu \nu}|_0 
& = & \left( \nabla_{{\bf E}_\nu} 
\nabla_{{\bf E}_\mu}
\langle {\bf E}_i, {\bf E}_j \rangle \right)|_0 \nonumber \\
& = & \left. \left \langle \nabla_{{\bf E}_i} { {\bf E}_\mu}, 
\nabla_{{\bf E}_j} {{\bf E}_\nu} \right \rangle\right|_0 
+ \left. \left \langle \nabla_{{\bf E}_i} {{\bf E}_\nu}, 
\nabla_{{\bf E}_j} {{\bf E}_\mu} \right \rangle \right|_0 \nonumber  \\
&& + \left. \left \langle \nabla_{{\bf E}_\nu} \nabla_{{\bf E}_i} {{\bf E}_\mu}, {\bf E}_j \right \rangle\right|_0 
+ \left. \left \langle {\bf E}_i , \nabla_{{\bf E}_\nu} \nabla_{{\bf E}_j} {{\bf E}_\mu} \right \rangle\right|_0.
\label{r26}
\end{eqnarray}
In the second equality, we again applied the Leibniz rule and
interchanged derivatives by virtue of Eqs.~(\ref{r8}) and (\ref{r6}).
We next note that the covariant derivative of ${\bf E}_\mu$ by ${\bf
E}_i$ is given by

\begin{equation}
\left. \left( \nabla_{{\bf E}_i} {{\bf E}_\mu} \right)\right|_0
= \left. \left( P_\parallel \nabla_{{\bf E}_i} 
{{\bf E}_\mu} 
+ P_\perp \nabla_{{\bf E}_i} 
{{\bf E}_\mu} \right)\right|_0
= \left. \left( T_{{\bf E}_i} {{\bf E}_\mu} 
+ S_{{\bf E}_i} {{\bf E}_\mu} \right) \right|_0, 
\label{r29}
\end{equation}
where the first equality follows from the fact that $P_\parallel +
P_\perp$ is the identity and the second from the definitions
Eqs.~(\ref{r25}) and (\ref{r27}) and the fact that ${\bf E}_\mu$ is
normal to ${\cal C}$.  We also observe from Eq.~(\ref{r28}) that
$\left (
\nabla_{{\bf E}_i} \nabla_{{\bf E}_\nu} {{\bf E}_\mu} \right)|_0 = 0$, and therefore

\begin{equation}
\left. \left( \nabla_{{\bf E}_\nu} \nabla_{{\bf E}_i} 
{{\bf E}_\mu}\right) \right|_0 = \left. \left( R_{{\bf E}_\nu {\bf E}_i} 
{{\bf E}_\mu}
\right)\right|_0,
\label{r30}
\end{equation}
where we have used Eqs.~(\ref{r6}) and (\ref{r31}).  Inserting
Eqs.~(\ref{r29}) and (\ref{r30}) into Eq.~(\ref{r26}) yields the desired result
Eq.~(\ref{r19}).

\section{The Constrained Hamiltonian}
\label{s3}

In Sect.~\ref{s11} we expanded the kinetic energy in $\epsilon$,
obtaining two terms.  One term, the transverse kinetic energy, is of
order $\epsilon^{-2}$; the other term is of order $\epsilon^0$.  In
this section, we apply (degenerate) first order perturbation theory to
derive a constrained Hamiltonian for the eigenenergies.  In doing so,
we introduce the transverse modes characterizing the wave function
away from the constraint manifold.

\subsection{Rescaling by $\epsilon$ and the expansion of the Hamiltonian}

By adding the potential energy $V_\perp({\sf u})$ to the kinetic energy
Eq.~(\ref{r32}), we have the following Hamiltonian

\begin{equation}
H =  H_\perp + H_\parallel + O(\epsilon),
\label{r39}
\end{equation}
where 

\begin{eqnarray}
H_\perp & = & K_\perp + V_\perp, 
\label{r130} \\
H_\parallel^p & = & K_\parallel^p + K_R + V_{ex}^p, 
\label{r104} 
\end{eqnarray}
are called the transverse and (preliminary) tangential Hamiltonians
respectively.

In order to clarify the subsequent perturbation analysis, we
explicitly exhibit the $\epsilon$ dependence of various quantities by
rescaling them in $\epsilon$.  To begin, we repeat the previous
definition Eq.~(\ref{r52}) of the rescaled quantities $\tilde{u}^\mu$
and also define rescaled momenta $\tilde{\pi}_\mu$,

\begin{eqnarray}
u^\mu 
& = & \epsilon \tilde{u}^\mu, 
\label{r53} \\
\pi_\mu 
& = & {1\over \epsilon} \tilde{\pi}_\mu.
\end{eqnarray}
Notice that both $\tilde{u}^\mu$ and $\tilde{\pi}_\mu$ scale as
$\epsilon^0$.  In general, the scaled version of a quantity (denoted
with a tilde) is defined such that the lowest order nonvanishing term
of its expansion in $\epsilon$ is of order $\epsilon^0$.  Thus, for a
quantity homogeneous in $\epsilon$, the scaled version is independent
of $\epsilon$.  For convenience, we repeat the definition
Eq.~(\ref{r54}) of the rescaled potential energy $\tilde{V}_\perp$ and
also define a rescaled transverse kinetic energy and transverse
Hamiltonian

\begin{eqnarray}
\tilde{V}_\perp (\tilde{\sf u}; \epsilon) 
& = & \epsilon^2 V_\perp({\sf u}; \epsilon) 
= \epsilon^2 V_\perp(\epsilon \tilde{\sf u}; \epsilon), \\
\tilde{K}_\perp & = & \epsilon^2 K_\perp
= {1 \over 2} \tilde{\pi}_\mu \tilde{\pi}_\mu, \\
\tilde{H}_\perp(\epsilon) 
& = & \epsilon^2 H_\perp (\epsilon) 
= \tilde{K}_\perp + \tilde{V}_\perp(\tilde{\sf u}; \epsilon). 
\label{r144}
\end{eqnarray}
By our previous assumptions in Sect.~\ref{s10},
$\tilde{V}_\perp(\tilde{\sf u};\epsilon)$ is smooth in $\epsilon$ and
does not vanish at $\epsilon = 0$.  As for $\tilde{K}_\perp$, it is
clearly independent of $\epsilon$.  Thus, $\tilde{H}_\perp(\epsilon)$
is smooth in $\epsilon$ at $\epsilon = 0$; its lowest order term is
order $\epsilon^0$, but depending on $\tilde{V}_\perp$, it may have
higher order terms as well.  Recall that $K_\parallel^p$, $K_R$,
$V_{ex}^p$, and $H_\parallel^p$ are already independent of $\epsilon$
and therefore need no further scaling.  For notational continuity,
however, we nevertheless define

\begin{eqnarray}
\tilde{K}_\parallel^p & = & K_\parallel^p, \\
\tilde{K}_R & = & K_R, \\
\tilde{V}_{ex}^p & = & V_{ex}^p, \\
\tilde{H}_\parallel^p & = & H_\parallel^p.
\end{eqnarray}

We  rescale the full Hamiltonian by defining

\begin{equation}
\tilde{H}(\epsilon)  
= \epsilon^2 H(\epsilon)
= \tilde{H}_\perp(\epsilon) 
+  \epsilon^2 \tilde{H}_\parallel^p
+ O(\epsilon^3).
\label{r38}
\end{equation}
In a typical Taylor series expansion of $\tilde{H}(\epsilon)$, we
would remove the order $\epsilon$ term from
$\tilde{H}_\perp(\epsilon)$ and leave it as a separate term.  We would
also combine the order $\epsilon^2$ term of
$\tilde{H}_\perp(\epsilon)$ with the tangential Hamiltonian
$\epsilon^2 \tilde{H}_\parallel^p$.  Here, however, we wish to keep the
$\epsilon$ and $\epsilon^2$ terms together in the transverse
Hamiltonian $\tilde{H}_\perp(\epsilon)$.  We therefore define a new
perturbation parameter $\kappa = \epsilon^2$ and rewrite
Eq.~(\ref{r38}) as

\begin{equation}
\tilde{H}(\epsilon, \kappa) 
= \tilde{H}_\perp(\epsilon) 
+ \kappa \tilde{H}_\parallel^p + O(\epsilon^3).
\label{r56}
\end{equation}
Our objective is to find the eigenvalues of $\tilde{H}$ through order
$\epsilon^2$.  Viewing $\epsilon$ and $\kappa$ as formally independent
in Eq.~(\ref{r56}), our objective becomes finding the eigenvalues of
$\tilde{H}$ through second order in $\epsilon$ and first order in
$\kappa$.  Our procedure is to assume that the eigenvalues of
$H_\perp(\epsilon)$ can be solved exactly (or at least through order
$\epsilon^2$) and then apply first order perturbation theory in
$\kappa$.  To simplify notation, we drop the $\epsilon$ dependence
(but not $\kappa$ dependence) for the duration of the derivation.

\subsection{Transformation to the Transverse Modes}

\label{s24}

The zeroth order term (in $\kappa$) of $\tilde{H}(\kappa)$ is the
transverse Hamiltonian $\tilde{H}_\perp$, which has the form

\begin{equation}
\tilde{H}_\perp 
= -{\hbar^2 \over 2} 
{\partial \over \partial \tilde{u}^\mu}
{\partial \over \partial \tilde{u}^\mu}
+ \tilde{V}_\perp(\tilde{\sf u}).
\end{equation}
Since $\tilde{H}_\perp$ depends only on the quantities
$\tilde{u}^\mu$, we may restrict its domain to functions of
$\tilde{u}^\mu$ alone.  For the moment we adopt this understanding for
the domain of $\tilde{H}_\perp$.  We pick an eigenvalue
$\tilde{E}_\perp$ (the transverse energy) of $\tilde{H}_\perp$ with
finite multiplicity $k$ and bounded eigenstates.  We call these
eigenstates the transverse modes (with energy $\tilde{E}_\perp$).  We
let $\chi_n(\tilde{\sf u})$, $n = 1, ...,k$, denote an orthonormal
basis of these transverse modes.  By orthonormal, we mean

\begin{equation}
\langle \chi_n | \chi_{n'} \rangle_{\sf u}
= \int du^1 \wedge ... \wedge du^d \;
\chi_n^*(\tilde{\sf u}) \chi_{n'}(\tilde{\sf u})
= \delta_{n n'},
\end{equation}
where the ${\sf u}$ subscript indicates integration only over the
variables $u^\mu$ as opposed to the full $(d + m)$-form $\nu$ in
Eq.~(\ref{r141}).

We now adopt the understanding that $\tilde{H}_\perp$ acts on wave
functions of both $\tilde{u}^\mu$ and $q$.  Such an eigenfunction with
eigenvalue $\tilde{E}_\perp$ has the general form

\begin{equation}
\psi(\tilde{\sf u}, q) 
= \sum_{n = 1}^k \chi_n(\tilde{\sf u}) \phi_n(q),
\label{r102}
\end{equation}
where the functions $\phi_n(q)$ are arbitrary.  We therefore identify
an eigenfunction $\psi(\tilde{\sf u}, q)$, having eigenvalue
$\tilde{E}_\perp$, with the $k$ functions $\phi_n(q)$.  Notice that
with our current understanding for the domain of the operator
$\tilde{H}_\perp$, $\tilde{E}_\perp$ is a degenerate eigenvalue, even
for $k = 1$.

Recall the steps involved in first order degenerate perturbation
theory.  First, one either proves or assumes that the desired
eigenvalue and eigenfunction is analytic in the perturbation parameter
$\kappa$.  (Here, we simply assume this fact.)  Next, one determines
the zeroth order energy and zeroth order eigenstates.  Then, one
considers the operator formed by restricting the first order term of
the Hamiltonian to the space of zeroth order eigenstates.  The first
order corrections to the energy are the eigenvalues of this restricted
operator.  In the present case, the zeroth order energy is
$\tilde{E}_\perp$, and the zeroth order eigenstates are given by
Eq.~(\ref{r102}).  The first order correction to the Hamiltonian is
$\kappa \tilde{H}_\parallel^p$.  Denoting the first order correction
to the energy by $\kappa
\tilde{E}_\parallel$, the eigenvalue equation for
$\tilde{E}_\parallel$ is

\begin{equation}
\sum_{n' = 1}^k \left(\tilde{H}_{\parallel}\right)_{ n n'} \phi_{n'} 
= \tilde{E}_\parallel  \phi_n,
\label{r103}
\end{equation}
where the $(\tilde{H}_{\parallel})_{n n'}$ are the differential operators 

\begin{equation}
\left(\tilde{H}_{\parallel}\right)_{ n n'} 
= \left \langle \chi_{n} \left| 
\tilde{H}_\parallel^p \chi_{n'} \right. \right \rangle_{\sf u} 
= \int d u^1 \wedge ... \wedge d u^d \chi_{n}^*(\tilde{\sf u})
\left(\tilde{H}_\parallel^p \chi_{n'}\right)(\tilde{\sf u}). 
\label{r134}
\end{equation}
We call $(\tilde{H}_\parallel)_{nn'}$ the constrained, or tangential,
Hamiltonian.

We now recall that $\kappa = \epsilon^2$ and reintroduce
the explicit $\epsilon$ dependence.  Summarizing our analysis thus
far, we have shown that an eigenvalue of $\tilde{H}(\epsilon)$ through
order $\epsilon^2$ is given by $\tilde{E}_\perp(\epsilon) + \epsilon^2
\tilde{E}_\parallel(\epsilon)$, where $\tilde{E}_\perp(\epsilon)$ and
$\tilde{E}_\parallel(\epsilon)$ are eigenvalues of
$\tilde{H}_\perp(\epsilon)$ and $(\tilde{H}_{\parallel})_{
n n'}(\epsilon)$.  Of course, assuming smoothness in $\epsilon$, it is sufficient to solve for
$\tilde{E}_\perp(\epsilon)$ and $\tilde{E}_\parallel(\epsilon)$
through orders $\epsilon^2$ and $\epsilon^0$ respectively.  We will
therefore only require $\tilde{E}_\parallel(\epsilon)$ and
$(\tilde{H}_{\parallel})_{ n n'}(\epsilon)$ evaluated at
$\epsilon = 0$, which we denote by $\tilde{E}_\parallel$ and
$(\tilde{H}_{\parallel})_{ n n'}$ respectively.  Also,
by virtue of Eq.~(\ref{r134}), we assume for the remainder of the
paper that the transverse modes $\chi_n(\tilde{\sf u})$ are only
order $\epsilon^0$ eigenfunctions of $\tilde{H}_\perp(\epsilon)$.

We view Eq.~(\ref{r103}) as a $k$-dimensional vector wave equation for
a vector wave function defined over the constraint manifold.  We
introduce the bold notation $\bbox{\phi}(q)$ for the vector wave
function with components $\phi_n(q)$ and the sans serif notation
$\tilde{\sf H}_\parallel$ for the matrix of differential operators
with components $(\tilde{H}_{\parallel})_{ nn'}$.
Equation~(\ref{r103}) can therefore be written more compactly as

\begin{equation}
{\sf H}_\parallel \bbox{\phi}
= {E}_\parallel  \bbox{\phi}.
\label{r129}
\end{equation}
Having completed the perturbation analysis, we have dropped the tildes
from ${\sf H}_\parallel$ and its eigenvalue $E_\parallel$.  Using
Eqs.~(\ref{r104}), (\ref{r55}), (\ref{r89}) and a little algebra,
we express ${\sf H}_\parallel$ as

\begin{equation}
{\sf H}_\parallel = {\sf K}_\parallel + {\sf V}_{ex},
\label{r105}
\end{equation}
where

\begin{eqnarray}
{\sf K}_\parallel 
& = & {1\over 2} 
(\pi^i|_0 {\sf I} 
+  S^{\mu \nu i}|_0 {\sf \Lambda}_{\mu \nu})^{\dagger} 
(\pi_i {\sf I} 
+  S^{\sigma \tau}_{\; \; \;\; i}|_0 {\sf \Lambda}_{\sigma \tau}), 
\label{r50} \\
{\sf V}_{ex} 
& = & V_{ex}^p {\sf I}
+ \left.\left( {1\over 2} S^{\mu \nu i } S^{\sigma \tau}_{\; \; \; \; i} 
+ {1 \over 6} R^{\mu \nu \sigma \tau} \right)\right|_0 
{\sf \Lambda}^{(2)}_{\mu \nu \sigma\tau} 
- \left. \left( {1\over 2} S^{\mu \nu i } S^{\sigma \tau}_{\; \; \; \; i}
\right)\right|_0 
 {\sf \Lambda}_{\mu \nu}  {\sf \Lambda}_{\sigma \tau}  \nonumber \\
& = &  V_{ex}^p {\sf I}
+ \left.\left( {1\over 2} S^{\mu \nu i} 
S^{\sigma \tau}_{\; \; \; \; i} \right)\right|_0
\left( {\sf \Lambda}^{(2)}_{\mu \nu \sigma \tau} 
- {\sf \Lambda}_{\mu \nu} {\sf \Lambda}_{\sigma \tau}\right)
+ {1 \over 6} R^{\mu \nu \sigma \tau}|_0 
{\sf \Lambda}^{(2)}_{\mu \nu \sigma\tau} 
\label{r51},
\end{eqnarray}
and where ${\sf I}$ is the $k \times k$ identity matrix and ${\sf
\Lambda}_{\mu \nu}$ and ${\sf \Lambda}^2_{\mu \nu \sigma\tau}$ are the
$k \times k$ matrices having the following components respectively

\begin{eqnarray}
(\Lambda_{\mu \nu})_{nn'} 
& = & \left \langle \chi_{n} \left| 
\Lambda_{\mu \nu} \chi_{n'} \right. \right \rangle_{\sf u},
\label{r42} \\
(\Lambda^{(2)}_{\mu \nu \sigma \tau})_{nn'} 
& = & \left \langle \chi_{n} 
\left| \Lambda_{\mu \nu} \Lambda_{\sigma \tau} 
\chi_{n'} \right. \right \rangle_{\sf u}.
\label{r131}
\end{eqnarray}
Equations~(\ref{r105}) -- (\ref{r51}) encapsulate the main result of
this paper.  Specifically, the constrained Hamiltonian ${\sf
H}_\parallel$ is a $k
\times k$ matrix of differential operators.  It is the residual
Hamiltonian remaining after the infinite transverse energy $E_\perp$
is subtracted off.  The kinetic energy ${\sf K}_\parallel$, which we
call the (final) tangential kinetic energy, differs from the
``standard'' kinetic energy in that it has a gauge potential term.
Physically, the gauge potential couples the tangential momenta to the
generalized angular momentum of the transverse modes.  The quantity
${\sf V}_{ex}$ is a $k \times k$ matrix of nondifferential operators
which we call the (final) extrapotential.  Notice that any possible
off-diagonal coupling in ${\sf H}_\parallel$ is due to the angular
momentum of the transverse modes.  The preliminary tangential kinetic
energy $K_\parallel^p$, extrapotential $V_{ex}^p$, and tangential
Hamiltonian $H_\parallel^p$ are distinguished from the (final)
tangential kinetic energy ${\sf K}_\parallel$, extrapotential ${\sf
V}_{ex}$, and tangential Hamiltonian ${\sf H}_\parallel$ by the
``$p$'' superscript.  We often drop the ``preliminary'' and ``final''
modifiers when referring to these terms, relying on their symbols and
context to make our precise meaning clear.

In the event of a nondegenerate transverse mode, that is $k=1$, ${\sf
H}_\parallel$ becomes a scalar wave operator $H_\parallel$ acting on
scalar wave functions defined over the constraint manifold.  In this
case, we see the emergence of Eqs.~(\ref{r79}) -- (\ref{r80})
presented in Sect.~\ref{s7}.  The exact derivation of these equations
from the more general Eqs.~(\ref{r105}) -- (\ref{r51}) will be
presented in Sect.~\ref{s12}.

\subsection{Nonconstant Transverse Potentials}

\label{s22}

Up to now, we have assumed that the transverse potential $V_\perp({\sf
u})$ is constant (modulo $SO(d)$ rotations) along the constraint
manifold ${\cal C}$.  For some physical systems this assumption holds
exactly due to some symmetry on the ambient space, such as $SO(3)$
rotations in the case of a rigid body.  However, for other systems,
this assumption may be only approximately satisfied; the constraining
potential may in fact vary along the constraint manifold.  This is
true, for example, of a molecule evolving along a reaction path; there
is no symmetry dictating that the frequencies of the small transverse
vibrations be constant.  The purpose of this section is to illustrate
how small variations in the transverse potential may be easily
included within our formalism.

The key idea is to only allow dependence on $q$ at order $\epsilon^2$.
Specifically, we assume the transverse potential can be expanded
as

\begin{equation}
\tilde{V}_\perp(\tilde{\sf u}, q; \epsilon)
= 
\tilde{V}_{\perp}^0(\tilde{\sf u})
+ \epsilon \tilde{V}_{\perp}^1(\tilde{\sf u})
+ \epsilon^2 \tilde{V}_{\perp}^2(\tilde{\sf u}, q)
+ O(\epsilon^3).
\end{equation}
Applying this expansion to Eq.~(\ref{r144}), an eigenvalue
$\tilde{E}_\perp$ of $\tilde{H}_\perp$ (assuming analyticity in
$\epsilon$) can be expanded as

\begin{equation}
\tilde{E}_\perp(q; \epsilon)
= 
\tilde{E}_{\perp}^0
+ \epsilon \tilde{E}_{\perp}^1
+ \epsilon^2 \tilde{E}_{\perp}^2(q)
+ O(\epsilon^3).
\end{equation}
The first two terms of $E_\perp = \tilde{E}_\perp/\epsilon^2$ blow up
as $\epsilon$ goes to $0$.  However, these two terms are constant in
$q$ and may thus be subtracted off.  The next order term
$\tilde{E}_{\perp}^2(q)$ does depend on $q$ and is of the same order
in $\epsilon$ as ${\sf H}_\parallel$.  Thus, $\tilde{E}_{\perp}^2(q)$
may be combined with the extrapotential ${\sf V}_{ex}$ in ${\sf
H}_\parallel$ to form the effective potential

\begin{equation}
{\sf V}_{ef}(q) = {\sf V}_{ex}(q) + \tilde{E}_{\perp}^2(q){\sf I}.
\end{equation}
This is the only modification which needs to be made to our formalism.
Notice that the transverse modes $\chi_n({\sf u})$ need not be
modified since they are defined to be only order $\epsilon^0$
eigenfunctions of $H_\perp$ and hence are unaffected by the term
$\tilde{E}_{\perp}^2(q)$.

\section{Analysis of Connections}

\label{s13}

Both the preliminary and the final tangential kinetic energies
$K_\parallel^p$ and ${\sf K}_\parallel$ exhibit a gauge potential
proportional to $S_{abc}$.  In this section, we study the geometric
origins of these gauge potentials and compute their curvatures. 

We begin by reviewing the connection on normal vector fields over
${\cal C}$.  We note that many equivalent definitions exist for the
general concept of a connection.  For the purposes of this paper, a
connection is taken to be a covariant derivative operator which acts
on some space of vector fields.  For more background, see any of a
number of standard references \cite{Kobayashi63,Spivak79b,Eguchi80}.
For the remainder of this section, ${\bf v}$ is an arbitrary normal
vector field over ${\cal C}$ and ${\bf x}$ and ${\bf y}$ are arbitrary
tangent vector fields over ${\cal C}$.  The normal connection
$\nabla^N$ is defined by

\begin{equation}
\nabla^N_{\bf x} {\bf v} 
= P_\perp \nabla_{\bf x} {\bf v} 
= P_\perp \nabla_{\bf x} P_\perp {\bf v},
\label{r33}
\end{equation}
where $\nabla$ is the Levi-Civita connection on ${\cal A}$.  Notice
that $\nabla^N_{\bf x} {\bf v}$ is itself a normal vector field.  The
curvature of $\nabla^N$, denoted $B^N$, is computed to be

\begin{eqnarray}
B^N_{{\bf x} {\bf y}} {\bf v} 
& = & \left( \nabla^N_{\bf x}  \nabla^N_{\bf y}
-  \nabla^N_{\bf y}  \nabla^N_{\bf x}
- \nabla^N_{[{\bf x}, {\bf y}]} \right) {\bf v} \nonumber \\
& = & P_\perp \left( \nabla_{\bf x} P_\perp \nabla_{\bf y}
-  \nabla_{\bf y} P_\perp \nabla_{\bf x}
- \nabla_{[{\bf x}, {\bf y}]} \right) P_\perp {\bf v} \nonumber \\
& = & P_\perp \left( R_{{\bf x} {\bf y}} 
- \nabla_{\bf x} P_\parallel \nabla_{\bf y} 
+ \nabla_{\bf y} P_\parallel \nabla_{\bf x} \right) P_\perp {\bf v}
=  P_\perp \left( R_{{\bf x} {\bf y}} 
- T_{\bf x} T_{\bf y} + T_{\bf y} T_{\bf x} \right) P_\perp {\bf v},
\label{r34}
\end{eqnarray}
where the first equality is simply the definition of the curvature,
the second follows from Eq.~(\ref{r33}), the third from noting
$P_\perp = I - P_\parallel$ and Eq.~(\ref{r31}), and the forth from
Eq.~(\ref{r25}). As expected, the curvature depends only on the nature
of the embedding of ${\cal C}$ (via the tensor $T$) and on the
curvature of ${\cal A}$.  If we assume that the tensors
$B^N$ and $R$ vanish, then we obtain the class of embeddings considered by
da Costa \cite{daCosta82}.  For such embeddings, one can choose a
potential frame with vanishing twist, thus eliminating coupling
between the transverse modes.  (This follows from Eq.~(\ref{r115}) below
and the fact that for vanishing curvature, one can always find a frame
for which the gauge potential vanishes.)  Also, for a non-twisting
potential frame, the submanifolds of constant potential are orthogonal to
the transverse spaces ${\cal U}_q$.  Hence, at all points the
restoring force is directed inward tangent to the ${\cal U}_q$.

It is instructive to compute the gauge potential explicitly for the
connection $\nabla^N$.  For this computation we first choose an
arbitrary orthonormal frame (not necessarily the potential frame)
${\bf V}_\mu$, $\mu = 1, ...,d$, for each normal space $N_q$.  We
denote the components of an arbitrary normal vector field ${\bf v}$
with respect to ${\bf V}_\mu$ by $v^\mu$.  Then, the components of
$\nabla^N_{{\bf E}_i} {\bf v}$ are given by

\begin{equation}
(\nabla^N_{{\bf E}_i} {\bf v})^\mu 
= {\bf E}_i v^\mu + (A_i^N)^\mu_{\;\; \nu} v^\nu,
\label{r35}
\end{equation}
where we have defined the gauge potential 

\begin{equation}
(A^N_i)_{\mu \nu} 
= \langle {\bf V}_\mu, \nabla^N_{{\bf E}_i} {\bf V}_\nu \rangle
= \langle {\bf V}_\mu, \nabla_{{\bf E}_i} {\bf V}_\nu \rangle.
\end{equation}
Due to the orthonormality of the ${\bf V}_\mu$, $(A^N_i)_{\mu \nu}$ is
antisymmetric in $\mu$ and $\nu$.  The gauge potential can therefore
be viewed as a one-form on ${\cal C}$ with values in the Lie algebra
$so(d)$, which contains all antisymmetric $d \times d$ matrices. If we
choose ${\bf V}_\mu = {\bf E}_\mu$, we recognize from Eq.~(\ref{r27})
that the gauge potential is related to the potential twist tensor by

\begin{equation}
(A^N_i)_{\mu \nu} = S_{\mu \nu i}.
\label{r115}
\end{equation}
This result will be import below for analyzing $K_\parallel^p$ and
${\sf K}_\parallel$.

We now consider a function $\psi({\bf v}, q)$, such as the quantum
wave function, defined in the neighborhood of ${\cal C}$.  (We use the
bold notation ${\bf v}$ instead of sans serif used earlier because we
wish to emphasize the dependence of $\psi$ on the normal vector and
not on its components with respect to a given frame, such as the
potential frame.)  The connection $\nabla^N$ which acts on normal
vector fields gives rise to another connection $\nabla^{\parallel p}$ which acts
on the function $\psi({\bf v}, q)$.  In order to define
$(\nabla^{\parallel p}_{\bf x}\psi)({\bf v}, q)$, we first choose a path
$q'(\alpha)$ such that $q'(0) = q$ and $(d q'/d
\alpha)(0) = {\bf x}$.  We then denote by ${\bf
v}'(\alpha)$ the unique normal vector at each point $q'(\alpha)$
satisfying ${\bf v}'(0) = {\bf v}$ and $(\nabla^N_{d/ d\alpha} {\bf
v}')(\alpha) = 0$.  Then, the connection $\nabla^{\parallel p}$ is defined by

\begin{equation}
(\nabla^{\parallel p}_{\bf x} \psi)({\bf v}, q) 
= \left. {d \over d\alpha}\right|_{\alpha = 0} 
\psi\bbox{(}{\bf v}'(\alpha), q'(\alpha)\bbox{)}.
\end{equation}

The transverse kinetic energy $K_\parallel^p$ can be directly related
to the covariant derivative $\nabla^{\parallel p}$.  To do this, it is
useful to compute the gauge potential of $\nabla^{\parallel p}$
explicitly.  As before, we consider an orthonormal frame ${\bf V}_\mu$
and denote the components of ${\bf v}$ by $v^\mu$.  Then, the function
$\psi({\bf v}, q)$ can also be interpreted as a function of $({\sf v},
q)$, where ${\sf v} = (v^1, ..., v^d)$ is the collection of
components.  We therefore have

\begin{eqnarray}
(\nabla^{\parallel p}_{{\bf E}_i} \psi)({\bf v}, q) 
& = & \left.{d \over d\alpha}\right|_{\alpha = 0} 
\psi\bbox{(}{\sf v}'(\alpha), q'(\alpha)\bbox{)}
=
\left.{d \over d\alpha}\right|_{\alpha = 0} 
\psi\bbox{(}{\sf v}'(\alpha), q\bbox{)} 
+ \left.{d \over d\alpha}\right|_{\alpha = 0} 
\psi\bbox{(}{\sf v}, q'(\alpha)\bbox{)} 
\nonumber \\
& = & 
({\bf E}_i {v'}^\mu)(q)
{\partial \psi \over \partial v^\mu}({\bf v}, q) 
+ ({\bf E}_i \psi)\bbox{(}{\bf v},q \bbox{)}, 
\label{r36}
\end{eqnarray}
where in the third equality, the derivatives $\partial / \partial
v^\mu$ and ${\bf E}_i$ are understood to have $q$ and $v^\nu$ held
fixed respectively.  From Eq.~(\ref{r35}) and the condition
$\nabla^N_{{\bf E}_i} {\bf v}' = 0$, we find

\begin{equation}
{\bf E}_i {v'}^\mu 
= -(A^N_i)^\mu_{\; \; \nu}{v'}^\nu.
\end{equation}
Inserting this result into Eq.~(\ref{r36}) yields

\begin{equation}
(\nabla^{\parallel p}_{{\bf E}_i} \psi)({\bf v}, q) 
= \left[ \bbox{(} {\bf E}_i 
+  A^{\parallel p}_i \bbox{)} \psi \right]({\bf v}, q),
\label{r37}
\end{equation}
where

\begin{equation}
A^{\parallel p}_i
= (A^N_i)^{\mu \nu} \Omega_{\mu \nu}, 
\label{r128}
\end{equation}
and where we have used the antisymmetry of $(A_i^N)_{\mu \nu}$ to
introduce the operator

\begin{equation}
\Omega_{\mu \nu} 
= {1\over 2}\left(v_\mu {\partial \over \partial v^\nu} -  v_\nu {\partial \over \partial v^\mu}\right).
\end{equation}
Obviously if ${\bf V}_\mu = {\bf E}_\mu$, then $\Lambda_{\mu \nu} =
-i\hbar {\Omega}_{\mu \nu}$. The relevance of $\nabla^{\parallel p}$
for $K_\parallel^p$ is now clear.  By choosing ${\bf V}_\mu = {\bf
E}_\mu$ and applying Eqs.~(\ref{r115}), (\ref{r37}), and (\ref{r128})
to Eq.~(\ref{r55}), we see that

\begin{equation}
K_\parallel^p 
= {\hbar^2 \over 2}(\nabla^{\parallel p}_{{\bf E}_i})^\dagger \nabla^{\parallel p}_{{\bf E}_i}.
\label{r117}
\end{equation}
Thus the preliminary tangential kinetic energy is just proportional to
the Laplacian defined in terms of the connection $\nabla^{\parallel
p}$.  (Compare Eq.~(\ref{r117}) to Eq.~(\ref{r73}).)

Considering Eq.~(\ref{r128}), we see that the two gauge potentials
$(A^N_i)_{\mu \nu}$ and $A^{\parallel p}_i$ differ only in their
representation of $so(d)$.  For $(A^N_i)_{\mu \nu}$, we use a
representation by $d\times d$ antisymmetric matrices, whereas for
$A^{\parallel p}_i$ we use a representation by the operators
$\Omega_{\mu \nu}$.  Therefore, the curvature of the connections
$\nabla^N$ and $\nabla^{\parallel p}$ are also related by simply
switching the representation of $so(d)$.  Hence the curvature $B^{\parallel p}$ of
$\nabla^{\parallel p}$ is

\begin{eqnarray}
B^{\parallel p}_{{\bf x} {\bf y}} \psi 
& = & \left( \nabla^{\parallel p}_{\bf x}  \nabla^{\parallel p}_{\bf y}
-  \nabla^{\parallel p}_{\bf y}  \nabla^{\parallel p}_{\bf x}
- \nabla^{\parallel p}_{[{\bf x}, {\bf y}]} \right) \psi
= (B^N_{{\bf x}{\bf y}})^{\mu \nu} \Omega_{\mu \nu} \psi \nonumber \\
& = & \left(R_{{\bf x} {\bf y}} 
- T_{\bf x} T_{\bf y} 
+ T_{\bf y} T_{\bf x} 
\right)^{\mu \nu} \Omega_{\mu \nu} \psi.
\label{r120}
\end{eqnarray}

We now consider a $k$-dimensional vector-valued function
$\bbox{\phi}(q)$ with components $\phi_n(q)$.  The connection
$\nabla^{\parallel p}$ induces a connection $\nabla^\parallel$ on
$\bbox{\phi}$ by the formula

\begin{equation}
(\nabla^\parallel_{\bf x} \bbox{\phi})_n
= \left \langle \chi_n \left| 
\nabla^{\parallel p}_{\bf x} 
\sum_{n' = 1}^k \chi_{n'} \phi_{n'} \right\rangle_{\sf u} \right. . 
\label{r116}
\end{equation}
The tangential kinetic energy ${\sf K}_\parallel$ is closely related
to the connection $\nabla^\parallel$ as we now show.  We take the
orthonormal frame ${\bf V}_\mu$ to be ${\bf E}_\mu$, and we recall
that $\chi_n({\sf u})$ is a function of $u^\mu$ alone and
$\bbox{\phi}(q)$ is a function of $q$ alone.  Then applying
Eqs.~(\ref{r115}), (\ref{r37}), and (\ref{r128}), we find

\begin{eqnarray}
\nabla^{\parallel p}_{\bf x}  \chi_{n} 
& = &  (S_{\bf x})^{\mu \nu} \Omega_{\mu \nu} \chi_{n}, 
\label{r119} \\
\nabla^{\parallel p}_{\bf x} \phi_n 
& = & {\bf x} \phi_n,
\end{eqnarray}
where $(S_{\bf x})_{\mu \nu} = \langle {\bf E}_\mu , S_{\bf x} {\bf
E}_\nu \rangle$.  Then  Eq.~(\ref{r116}) yields

\begin{eqnarray}
\nabla^\parallel_{\bf x} 
& = & {\bf x}{\sf I} + {\sf A}_{\bf x}^\parallel, 
\label{r118} \\
{\sf A}_{\bf x}^\parallel
& = & (S_{\bf x})^{\mu \nu} {\sf \Omega}_{\mu \nu},
\label{r126}
\end{eqnarray}
where ${\sf \Omega}_{\mu \nu}$ is the $k \times k$ matrix with components

\begin{equation}
(\Omega_{\mu \nu})_{nn'} = \langle \chi_n 
| \Omega_{\mu \nu} \chi_{n'} \rangle_{\sf u}.
\end{equation}
From Eq.~(\ref{r119}), we note that the components of ${\sf A}_{\bf
x}^\parallel$ can also be written as

\begin{equation}
({A}_{\bf x}^\parallel)_{nn'}
= \langle \chi_n | \nabla^{\parallel p}_{\bf x} \chi_{n'} \rangle_{\sf u}.
\label{r127}
\end{equation}
Equations~(\ref{r118}) and (\ref{r126}) show that the tangential
kinetic energy ${\sf K}_\parallel$, Eq.~(\ref{r50}), is given by

\begin{equation}
{\sf K}_\parallel 
= {\hbar^2 \over 2} (\nabla_{{\bf E}_i}^\parallel)^\dagger
\nabla_{{\bf E}_i}^\parallel,
\label{r133}
\end{equation}
analogous to Eq.~(\ref{r117}) for $K_\parallel^p$.

The connection $\nabla^\parallel$ is closely related to the adiabatic
transport of quantum states and the associated geometric phase due to
Berry~\cite{Berry84}.  If a set of $k$ degenerate quantum states
$\xi_n({\sf \eta})$, $n = 1, ..., k$, depending smoothly on a set of
$m$ external parameters ${\sf
\eta} = (\eta_1, ..., \eta_m)$, is subject to an adiabatic variation
${\sf \eta}(\alpha)$ of these parameters, then the $\xi_n(\alpha) =
\xi_n({\sf \eta}(\alpha))$ satisfy $\langle \xi_n | d
\xi_{n'} / d\alpha \rangle = \sum_i
\langle
\xi_n | \partial \xi_{n'} /
\partial \eta_i \rangle (d\eta_i/d\alpha) = 0$.  Simon \cite{Simon83} recognized that this condition defines a connection 

\begin{equation}
\nabla^B_{\partial / \partial \eta_i} 
= {\partial \over \partial \eta_i}{\sf I} + {\sf A}^B_i
\label{r125}
\end{equation}
acting on the vector-valued wave function $\bbox{\xi} = (\xi_1, ..., \xi_k)$
parameterized by ${\sf \eta}$.  The gauge potential ${\sf A}^B_i$ is a
$k \times k$ matrix with components

\begin{equation}
({A}_i^B)_{nn'} 
= \left \langle \xi_n \left 
| {\partial \xi_{n'} \over \partial \eta_i} \right \rangle\right..  
\label{r142}
\end{equation}
If the parameters $\eta_i$ are themselves quantized, then the
momentum conjugate to $\eta_i$ is not simply $-i\hbar(\partial /
\partial \eta_i) {\sf I}$, but rather $-i\hbar
\nabla^B_{\partial / \partial \eta_i} = -i\hbar({\partial / \partial
\eta_i} {\sf I} + {\sf A}^B_i)$.  This situation applies, for example, to the
Born-Oppenheimer theory of molecules, wherein the parameters $\eta_i$
describe the positions of the nuclei and the $\xi_n$
represent the quantum state of the electrons\cite{Mead92}.  For the
constrained quantum systems considered in this paper, the ordering in
$\epsilon$ adiabatically separates the transverse modes $\chi_n$
(analogous to the $\xi_n$) from the motion along the
constraint manifold (analogous to the space of $\eta_i$).  Therefore,
the gauge potential ${\sf A}^\parallel_{\bf x}$ occurring in
Eq.~(\ref{r118}) is essentially the same as Berry's gauge potential
${\sf A}^B_i$ occurring in Eq.~(\ref{r125}).  We say ``essentially the
same'' because the coordinate derivative $\partial / \partial \eta_i$
of Eq.~(\ref{r142}) has been replaced by the covariant derivative
$\nabla^{\parallel p}$ of Eq.~(\ref{r127}), this covariant derivative
being the geometrically natural connection for the transverse modes.

We next compute the curvature of the connection $\nabla^\parallel$.
In terms of the gauge potential ${\sf A}^\parallel = (S_{\bf x})^{\mu
\nu} {\sf \Omega}_{\mu \nu}$, we have

\begin{equation}
{\sf B}^\parallel_{{\bf x} {\bf y}}
= (d {\sf A}^\parallel)({\bf x}, {\bf y}) 
+ [ {\sf A}_{\bf x}^\parallel, {\sf A}_{\bf y}^\parallel ]
= (d S^{\mu \nu})({\bf x}, {\bf y}) {\sf \Omega}_{\mu \nu}
+ (S_{\bf x})^{\mu \nu} (S_{\bf y})^{\sigma\tau} 
[ {\sf \Omega}_{\mu \nu}, {\sf \Omega}_{\sigma \tau} ],
\label{r146}
\end{equation}
where $d S^{\mu \nu}$ is the exterior derivative of $S^{\mu \nu}$,
viewed as a one-form over ${\cal C}$.  We determine $d S^{\mu
\nu}$ from the formula Eq.~(\ref{r120}) for the curvature $B^{\parallel
p}$.  We first note

\begin{equation}
{B}^{\parallel p}_{{\bf x} {\bf y}}
= (d {A}^{\parallel p})({\bf x}, {\bf y}) 
+ [ {A}_{\bf x}^{\parallel p}, {A}_{\bf y}^{\parallel p} ]
= (d S^{\mu \nu})({\bf x}, {\bf y}) {\Omega}_{\mu \nu}
+ (S_{\bf x})^{\mu \nu} (S_{\bf y})^{\sigma\tau}  
[ {\Omega}_{\mu \nu}, {\Omega}_{\sigma \tau} ],
\label{r145}
\end{equation}
where we have used Eqs.~(\ref{r115}) and (\ref{r128}).  It is
straightforward to verify that the ${\Omega}_{\mu \nu}$ satisfy the
following commutation relations,

\begin{equation}
[ {\Omega}_{\mu \nu}, {\Omega}_{\sigma \tau} ]
= {1 \over 2}(\delta_{\mu \sigma} {\Omega_{\tau \nu}} + 
\delta_{\nu \tau} {\Omega_{\sigma \mu}} +
\delta_{\mu \tau} {\Omega_{\nu \sigma}} +
\delta_{\nu \sigma} {\Omega}_{\mu \tau}),
\label{r148}
\end{equation}
and hence Eq.~(\ref{r145}) reduces to

\begin{equation}
{B}^{\parallel p}_{{\bf x} {\bf y}}
= [(d S^{\mu \nu})({\bf x}, {\bf y}) 
+ (S_{\bf x} S_{\bf y} - S_{\bf y} S_{\bf x})^{\mu \nu}] 
{\Omega}_{\mu \nu}.
\end{equation}
Combining this equation with Eq.~(\ref{r120}) produces 

\begin{equation}
(d S^{\mu \nu})({\bf x}, {\bf y}) 
= ( R_{{\bf x}{\bf y}} - T_{\bf x} T_{\bf y} + T_{\bf y} T_{\bf x}
- S_{\bf x} S_{\bf y} + S_{\bf y} S_{\bf x} )^{\mu \nu}.
\label{r147}
\end{equation}
Combining Eq.~(\ref{r147}) in turn with Eq.~(\ref{r146}), we arrive at
the following useful formula for the curvature of $\nabla^\parallel$,

\begin{equation}
{\sf B}^\parallel_{{\bf x} {\bf y}}
= ( R_{{\bf x}{\bf y}} - T_{\bf x} T_{\bf y} + T_{\bf y} T_{\bf x}
- S_{\bf x} S_{\bf y} + S_{\bf y} S_{\bf x} )^{\mu \nu}{\sf \Omega}_{\mu \nu}
+ (S_{\bf x})^{\mu \nu} (S_{\bf y})^{\sigma \tau} 
[ {\sf \Omega}_{\mu \nu}, {\sf \Omega}_{\sigma \tau} ].
\end{equation}
Using the commutation relations Eq.~(\ref{r148}) the above equation can be
recast as

\begin{eqnarray}
B^\parallel_{{\bf x} {\bf y}}
&=& \left(R_{{\bf x} {\bf y}} 
- T_{\bf x} T_{\bf y} 
+ T_{\bf y} T_{\bf x} 
\right)^{\mu \nu} {\sf \Omega}_{\mu \nu}  \nonumber \\
&& + [(S_{\bf x})^{\mu \nu}(S_{\bf y})^{\sigma \tau}
- (S_{\bf y})^{\mu \nu}(S_{\bf x})^{\sigma \tau}] 
({\sf \Omega}_{\mu \nu} {\sf \Omega}_{\sigma \tau}
 - {\sf \Omega}^{(2)}_{\mu \nu \sigma \tau}),
\label{r132}
\end{eqnarray}
where ${\sf \Omega}^{(2)}_{\mu \nu \sigma \tau}$ is the $k \times k$ matrix with components

\begin{equation}
({\Omega}^{(2)}_{\mu \nu \sigma \tau})_{nn'} 
= \langle \chi_n| 
\Omega_{\mu \nu} \Omega_{\sigma \tau} \chi_{n'} \rangle.
\end{equation}

\section{Specific Cases and Examples}

\label{s17}

We consider several concrete examples to help clarify the general
theory. 

\subsection{Codimension One Case}

We assume here that the codimension of the constraint manifold is $d =
1$.  Since there is only one normal direction, we expect the potential
twist to vanish.  Indeed, this follows from the antisymmetry property
$S_{\mu \nu i} = -S_{\nu \mu i}$ (Eq.~(\ref{r114})) and the fact that
$\mu = \nu =1$.  Similarly, the normal components of the Riemannian
curvature $R_{\mu \nu
\sigma \tau}$ also vanish due to the well-known antisymmetry
property $R_{abcd} = - R_{bacd} = - R_{abdc}$.  From this fact follows
${\cal R}_\perp = R^{11}_{\; \; \; \; 11}|_0 = 0$ and ${\cal R} =
{\cal R}_\parallel$.  The expressions for ${\cal T}^2$ and ${\cal
M}^2$ can also be simplified by introducing the rank two symmetric
tensor $W$ defined on vectors tangent to ${\cal C}$ and with
components $W^i_{\; \; j} = T^i_{\; \; 1 j}$.  (This tensor is often
called the Weingarten map.)  Then ${\cal T}^2 =
\mbox{Tr}\; (W^2)$ and ${\cal M}^2 = (\mbox{Tr}\; W)^2$.  Hence,
the tangential Hamiltonian Eq.~(\ref{r105}) becomes

\begin{eqnarray}
{\sf H}_\parallel 
& = & {\sf K}_\parallel + {\sf V}_{ex}, \\
{\sf K}_\parallel 
& = & {1\over 2} {\pi^i|_0}^\dagger \pi_i {\sf I}, \\
{\sf V}_{ex} 
& = & V_{ex}^p {\sf I}
= -{\hbar^2 \over 8} \left( 
{\cal T}^2 - \hat{\cal R} + {\cal R}_\parallel \right) {\sf I} \nonumber \\
& = & -{\hbar^2 \over 8} \left( 
{\cal M}^2 - 2\hat{\cal R} + 2 {\cal R}_\parallel \right) {\sf I} 
=  -{\hbar^2 \over 8} \left( 
2 {\cal T}^2 - {\cal M}^2 \right) {\sf I},
\label{r106}
\end{eqnarray}
where we have used Eqs.~(\ref{r143}), (\ref{r107}), and (\ref{r135}).
Notice that the tangential kinetic energy is proportional to the
standard Laplacian on ${\cal C}$.  All reference to $\Lambda_{\mu
\nu}$ has vanished, and hence all coupling between the degenerate
transverse modes has been eliminated.  The $k$-dimensional
Schr\"{o}dinger equation therefore separates into $k$ independent
scalar Schr\"{o}dinger equations.

We consider the case where the ambient space ${\cal A}$ is a flat
two-dimensional space and the constraint manifold ${\cal C}$ is a
curve in that space.  Then, we note that $\hat{\cal R} = {\cal R} =
{\cal R}_\parallel = 0$.  Furthermore, the second fundamental form, or
equivalently the Weingarten map, has only one nonzero component. We
denote this component by $W = W_{ii} =
\kappa = 1 /
\rho$, where $\kappa$ is the external curvature and $\rho$ is the
radius of curvature.  Then, the extrapotential is

\begin{equation}
V_{ex}^p 
= - {\hbar^2 \over 8} {1 \over \rho^2}
= - {\hbar^2 \over 8} {\kappa}^2.
\end{equation}
As in Sect.~\ref{s7} the sign on $V_{ex}^p$ is such that $\phi$ is
attracted to regions of high curvature.  This extrapotential was
derived earlier by Marcus~\cite{Marcus66} and Switkes, Russell, and Skinner \cite{Switkes77}.

We next consider the case where ${\cal A}$ is a flat three-dimensional
space and ${\cal C}$ is a two-dimensional surface.  We still have that
${\cal R} = {\cal R}_\parallel = 0$.  Furthermore, the eigenvalues of the
second rank two-dimensional tensor $W_{ij}$ are $\kappa_1 = 1/\rho_1$ and
$\kappa_2 = 1/ \rho_2$, where $\rho_1$ and $\rho_2$ are the two
external radii of curvature.  Then the extrapotential $V_{ex}^p$ is
conveniently written

\begin{equation}
V_{ex}^p 
= - {\hbar^2 \over 8} \left[ 2 \mbox{Tr}\; (W^2) - (\mbox{Tr}\; W)^2 \right]
= - {\hbar^2 \over 8} \left({1\over \rho_1} - {1 \over \rho_2}\right)^2
= - {\hbar^2 \over 8} \left( \kappa_1 - \kappa_2\right)^2.
\end{equation}
This result was previously derived by Jensen and Koppe \cite{Jensen71}
as well as da Costa \cite{daCosta81}.

\subsection{Codimension Two Case}
\label{s12}

We assume here that the codimension of the constraint manifold is $d =
2$.  This allows us to define the quantities $S_i$, ${\sf
\Lambda}$, and ${\sf \Lambda}^{(2)}$  by

\begin{eqnarray}
S_{\mu \nu i} 
& = & S_i \epsilon_{\mu \nu}, \\
{\sf \Lambda}_{\mu \nu}
& = & {\sf \Lambda} \epsilon_{\mu \nu}, \\
{\sf \Lambda}^{(2)}_{\mu \nu \sigma \tau}
& = & {\sf \Lambda}^{(2)} \epsilon_{\mu \nu} \epsilon_{\sigma \tau},
\end{eqnarray}
where $\epsilon_{\mu \nu}$ is the $2 \times 2$ antisymmetric tensor
with $\epsilon_{12} = -\epsilon_{21} = 1$.  Furthermore, we have

\begin{equation}
R_{\mu \nu \sigma \tau} 
= {1\over 2}{\cal R}_\perp \epsilon_{\mu \nu} \epsilon_{\sigma \tau}.
\end{equation}
We express the tangential Hamiltonian Eq.~(\ref{r105}) as

\begin{eqnarray}
{\sf H}_\parallel 
& = & {\sf K}_\parallel + {\sf V}_{ex}, \\
{\sf K}_\parallel 
& = & {1\over 2} (\pi^{i}|_0{\sf I} + 2 S^i|_0 {\sf \Lambda})^\dagger
 (\pi_i{\sf I} + 2 S_i {\sf \Lambda}), 
\label{r108} \\
{\sf V}_{ex} 
& = & V_{ex}^p {\sf I} 
+ (2 S^i S_i)|_0 ({\sf \Lambda}^{(2)} - {\sf \Lambda}^2)
+ {1 \over 3} {\cal R}_\perp|_0 {\sf \Lambda}^2.
\label{r111}
\end{eqnarray}

We consider the case of Sect.~\ref{s7} where ${\cal A}$ is a flat
three-dimensional space and ${\cal C}$ is a one-dimensional curve.
First, we note $\hat{\cal R} = {\cal R} = {\cal R}_\perp = {\cal
R}_\parallel = 0$.  Next, we denote the single component of tangential
momentum by $\pi_\parallel =
\pi_i$.  Since ${\cal C}$ is one dimensional, $\pi_\parallel = -i\hbar
\partial / \partial \alpha = \pi_\parallel^\dagger$, where $\alpha$
is the geodesic length.  Furthermore, the potential twist is
determined by the sole component ${\cal S} = S_i$.  The second
fundamental form can be identified with a normal vector $T^\mu =
T^{i\mu i}$ of magnitude $\kappa = 1/\rho$.  Hence, ${\cal M}^2 = T^{i
\mu}_{\;\; \; \; i} T^j_{\; \; \mu j} = T^\mu T_\mu = \kappa^2 = 1/
\rho^2$.  Using Eq.~(\ref{r135}), Eqs.~(\ref{r108}) and (\ref{r111})
therefore simplify to

\begin{eqnarray}
{\sf K}_\parallel 
& = & {1\over 2} (\pi_\parallel {\sf I} + 2 {\cal S} {\sf \Lambda})^2,
\label{r112} \\
{\sf V}_{ex} 
& = & -{\hbar^2 \over 8} {\kappa}^2 {\sf I} 
+ (2 {\cal S}^2) ({\sf \Lambda}^{(2)} - {\sf \Lambda}^2).
\label{r113} 
\end{eqnarray}
Assuming a single nondegenerate transverse mode, Eqs.~(\ref{r112})
and (\ref{r113}) yield Eqs.~(\ref{r110}) and (\ref{r80}). 

\subsection{Rotationally Invariant Transverse Potential}

In this section, we assume the transverse potential $V_\perp({\sf u})$
is rotationally invariant, depending only on the radius $u = (u^\mu
u^\mu)^{1/2}$ in the normal space.  The potential frame ${\bf E}_\mu$
can therefore be any orthonormal frame we like.  This freedom in the
choice of potential frame produces a large range of possible potential
twist tensors $S$, with the actual choice of $S$ being simply a matter
of convention.  The Hamiltonian ${\sf H}_\parallel$ in
Eq.~(\ref{r105}), however, should be independent (up to a rescaling of
the wave function $\bbox{\phi}$) of any such conventions.  In the
remainder of this section, we show explicitly how the dependence on
$S$ drops out of ${\sf H}_\parallel$ under the assumption of
rotational invariance.

First, we observe that the transverse Hamiltonian Eq.~(\ref{r130}) has
the form

\begin{equation}
H_\perp 
= -{\hbar^2 \over 2} {1 \over u^{d-1}} {\partial \over \partial u} 
u^{d-1} {\partial \over \partial u}
+ {\Lambda^2 \over 2 u^2} + V_\perp(u),
\end{equation}
where $\Lambda^2$ is the Casimir operator

\begin{equation}
\Lambda^2 = \Lambda_{\mu \nu} \Lambda^{\mu \nu}.
\end{equation}
Therefore, an eigenfunction $\chi_n$ of $H_\perp$ is necessarily an
eigenfunction of $\Lambda^2$.  We denote by $\chi^\lambda_n$ such an
eigenfunction whose $\Lambda^2$ eigenvalue is $\lambda$.  A basic fact
concerning the eigenspaces of the Casimir $\Lambda^2$ is that they
block diagonalize the generators $\Lambda_{\mu \nu}$.  That is,
$\langle \chi_n^\lambda |
\Lambda_{\mu \nu} \chi^{\lambda'}_{n'} \rangle_{\sf u} = 0$ if $\lambda \ne
\lambda'$.\footnote{This follows quickly from  $[\Lambda^2, \Lambda_{\mu \nu}] = 0$.  Note, $(\lambda - \lambda') \langle \chi_n^\lambda |
\Lambda_{\mu \nu} \chi^{\lambda'}_{n'} \rangle_{\sf u}
= \langle \chi_n^\lambda | [ \Lambda^2, \Lambda_{\mu \nu} ] \chi_{n'}^{\lambda'}
\rangle_{\sf u} = 0$.}
Based on the definitions Eqs.~(\ref{r42}) and (\ref{r131}) for ${\sf
\Lambda}^{(2)}_{\mu \nu \sigma \tau} $ and ${\sf \Lambda}_{\mu
\nu}$, this fact implies that for the space of transverse modes for a given $E_\perp$, 
${\sf \Lambda}^{(2)}_{\mu \nu \sigma \tau} = {\sf \Lambda}_{\mu \nu}
{\sf \Lambda}_{\sigma \tau}$.  Similarly, ${\sf \Omega}^{(2)}_{\mu \nu
\sigma
\tau} = {\sf \Omega}_{\mu \nu} {\sf \Omega}_{\sigma \tau}$.  We therefore see from Eq.~(\ref{r51}) that all $S$ dependence drops out of ${\sf V}_{ex}$,

\begin{equation}
{\sf V}_{ex} 
=   V_{ex}^p {\sf I}
+ {1 \over 6} R^{\mu \nu \sigma \tau}|_0 
{\sf \Lambda}_{\mu \nu}{\sf \Lambda}_{\sigma\tau}. 
\end{equation}

Considering ${\sf K}_\parallel$, even though Eq.~(\ref{r50}) is
written in terms of the potential twist $S$, we showed in
Sect.~\ref{s13} (specifically Eq.~(\ref{r133})) that ${\sf
K}_\parallel$ can be expressed in terms of the Laplacian associated
with the connection $\nabla^\parallel$.  From Eq.~(\ref{r132}) and the
results above, we see that the curvature $B^\parallel$ of this
connection is independent of $S$,

\begin{equation}
B^\parallel_{{\bf x} {\bf y}}
= \left(R_{{\bf x} {\bf y}} 
- T_{\bf x} T_{\bf y} 
+ T_{\bf y} T_{\bf x} 
\right)^{\mu \nu} {\sf \Omega}_{\mu \nu}.  
\end{equation}
Now if two connections $\nabla^\parallel$ and ${\nabla^\parallel}'$
have the same curvature, then their associated Laplacians can only
differ by a rescaling of the wave function.  Hence, the Hamiltonian
${\sf H}_\parallel$ for different choices of the potential twist $S$ can
at most differ by such a rescaling.  

\subsection{Harmonic Transverse Potentials}
\label{s21}

We assume that the transverse potential is quadratic in the $u^\mu$

\begin{equation}
V_\perp({\sf u}; \epsilon) 
= \sum_\mu {1\over 2} (\omega_\mu(\epsilon))^2 u^\mu u^\mu
\label{r150}
\end{equation}
and that the oscillation frequencies depend on $\epsilon$ via
$\omega_\mu(\epsilon) = \tilde{\omega}_\mu/\epsilon^2$, with
$\tilde{\omega}_\mu$ being independent of $\epsilon$.  (For clarity,
we make summation over the indices $\mu$, $\nu$, $\sigma$,
... explicit in this section.)  We introduce the standard machinery of
raising, lowering, and number operators for each degree of freedom,

\begin{eqnarray}
a_\mu 
& = & {1\over \sqrt{2 \hbar}} 
\left( \sqrt{\omega_\mu} u_\mu 
+ i {\pi_\mu \over \sqrt{\omega_\mu}} \right), \\
u^\mu 
& = & \sqrt{\hbar \over 2 \omega_\mu} \left(a_\mu + a_\mu^\dagger \right), 
\label{r44} \\
\pi_\mu 
& = & -i \sqrt{\hbar \omega_\mu \over 2} \left(a_\mu - a_\mu^\dagger \right), 
\label{r45} \\
N_\mu & = & a_\mu^\dagger a_\mu, \\
\left[ a_\mu, a_\nu^\dagger \right] & = & \delta_{\mu \nu}.
\end{eqnarray}
Notice that $a_\mu$ and $N_\mu$ scale as $\epsilon^0$.  The transverse
Hamiltonians $H_\perp$ and $\tilde{H}_\perp$  have the usual form

\begin{eqnarray}
H_\perp(\epsilon) 
& = & \sum_\mu \hbar \omega_\mu(\epsilon) \left( N_\mu + {1\over 2} \right), \\
\tilde{H}_\perp 
& = & \sum_\mu \hbar \tilde{\omega}_\mu \left(N_\mu + {1\over 2} \right),
\end{eqnarray} 
and the transverse modes can therefore be labeled by the number of
quanta $n_\mu$ in each degree of freedom $\mu$.  We denote such a mode
by $\chi_{\sf n}$ where ${\sf n} = (n_1, ..., n_d)$.  Inserting
Eqs.~(\ref{r44}) and (\ref{r45}) into Eq.~(\ref{r46}) yields

\begin{equation}
\Lambda_{\mu \nu} 
= {i \hbar \over 4 \sqrt{\omega_\mu \omega_\nu}} 
\left[ (\omega_\mu - \omega_\nu) (a_\mu  a_\nu - a_\mu^\dagger a_\nu^\dagger) 
+ (\omega_\mu + \omega_\nu) ( a_\nu^\dagger a_\mu - a_\mu^\dagger a_\nu ) \right],
\label{r47}
\end{equation}
from which one quickly sees

\begin{equation}
\left \langle \chi_{\sf n} 
| \Lambda_{\mu \nu} \chi_{\sf n} \right \rangle_{\sf u} 
= 0.
\label{r48}
\end{equation}
A significantly more involved computation yields

\begin{equation}
\langle \chi_{\sf n} | \Lambda_{\mu \nu} 
\Lambda_{\sigma \tau} \chi_{\sf n} \rangle_{\sf u}
= 
- {\hbar^2 \over 8} \left[
2 \left(n_\mu + {1\over 2} \right) \left(n_\nu + {1 \over 2}\right) 
{\omega_\mu^2 + \omega_\nu^2 \over \omega_\mu \omega_\nu} - 1 
\right]
(\delta_{\mu \tau} \delta_{\nu \sigma} - \delta_{\mu \sigma} \delta_{\nu \tau} ).
\label{r49}
\end{equation}

We now assume that $\chi = \chi_{\sf n}$ is a nondegenerate transverse
mode.  The tangential Hamiltonian Eq.~(\ref{r105}) is therefore a
scalar operator.  Using Eqs.~(\ref{r48}) and (\ref{r49}),
$H_\parallel$ is

\begin{eqnarray}
H_\parallel & = & K_\parallel + V_{ex}, \\
K_\parallel & = & {1\over 2} {\pi^i|_0}^\dagger \pi_i, 
\label{r149} \\
V_{ex}
& = & V_{ex}^p 
- {\hbar^2 \over 8} \sum_{\mu \nu} \left. \left( S^{\mu \nu i} S_{\mu \nu i} 
+ {1 \over 3} R^{\mu \nu}_{\; \; \; \; \mu \nu} \right)\right|_0 
\left[1 - 2 \left(n_\mu + \frac{1}{2} \right) \left(n_\nu + {1 \over 2}\right) 
{\omega_\mu^2 + \omega_\nu^2 \over \omega_\mu \omega_\nu}\right].
\label{r57}
\end{eqnarray}
The most striking aspect of the above equations is that, due to the
vanishing of $\langle \chi| {\Lambda}_{\mu \nu} \chi \rangle_{\sf u}$, the
tangential kinetic energy $K_\parallel$ is proportional to the
standard Laplacian on the constraint manifold.  Thus, all of the
effects of external curvature and potential twist are contained in the
extrapotential $V_{ex}$.

\subsection{Potentials with Reflection Symmetry}

\label{s23}

The vanishing of $\langle \chi |{\Lambda}_{\mu \nu} \chi \rangle_{\sf
u}$ (and hence the potential twist as well) from ${K}_\parallel$ in
Eq.~(\ref{r149}) follows from general considerations of reflection
symmetry, and therefore occurs for a large class of symmetric
potentials. 

Let ${\sf Q} \in O(d)$ be a reflection acting on the transverse
coordinates ${\sf u} = (u^1 , ..., u^d)$, and assume that, for a given
$\sigma$, $u^\sigma$ is mapped to $-u^\sigma$ and all other
coordinates remain fixed.  Thus, ${\sf Q} = {\sf Q}^{-1} = {\sf
Q}^\dagger$.  Furthermore, assume that $V_\perp({\sf u})$ is invariant
under the action of ${\sf Q}$, that is $V_\perp({\sf Q}{\sf u}) =
V_\perp({\sf u})$.  The reflection ${\sf Q}$ also has an induced
action on the transverse modes, which we denote by $Q$ and which is
given by $(Q
\chi_n)({\sf u}) = \chi_n( {\sf Q}^{-1} {\sf u})$.  Due to the symmetry of
$V_\perp$, $Q$ commutes with $H_\perp$,

\begin{equation}
[Q, H_\perp] = 0.
\label{r151}
\end{equation}
Furthermore, the following are easily verified

\begin{eqnarray}
Q u^\sigma Q^\dagger 
& = & -u^\sigma, \\
Q \pi_\sigma Q^\dagger 
& = & -\pi_\sigma, \\
Q \Lambda_{\sigma \mu} Q^\dagger 
& = & -\Lambda_{\sigma \mu} \hskip 1cm \mbox{for all $\mu$}.
\label{r152}
\end{eqnarray}

We now consider a single nondegenerate transverse mode, denoted simply
by $\chi$.  Due to Eq.~(\ref{r151}), $\chi$ must also be an
eigenfunction of $Q$ with eigenvalue either $+1$ or $-1$ (since $Q^2 =
I$).  Combining these facts with Eq.~(\ref{r152}) and recalling $Q =
Q^\dagger$, we have

\begin{equation}
\langle \chi | \Lambda_{\sigma \mu} \chi \rangle_{\sf u}
= \langle Q \chi | \Lambda_{\sigma \mu} Q \chi \rangle_{\sf u}
= \langle \chi_n | Q \Lambda_{\sigma \mu} Q^\dagger \chi_n \rangle_{\sf u}
= - \langle \chi | \Lambda_{\sigma \mu} \chi \rangle_{\sf u},
\end{equation}
and hence

\begin{equation}
\langle \chi | \Lambda_{\sigma \mu} \chi \rangle_{\sf u} = 0
\hskip 1cm \mbox{for all $\mu$}.
\end{equation}

If the potential $V_\perp({\sf u})$ is symmetric with respect to at
least $d-1$ such ${\sf Q}$ reflections, possessing $d-1$ distinct and
orthogonal reflection axes $u^\sigma$, then $\langle \chi |
\Lambda_{\sigma \mu} \chi \rangle_{\sf u}$ vanishes for all $\mu, \nu = 1,
..., d$.  For such highly symmetric potentials, $K_\parallel$ is again
given by Eq.~(\ref{r149}) and the only effect of the potential twist
is to be found in $V_{ex}$.  This is the case for such common
potentials as the simple harmonic oscillator, analyzed in the last
section, as well as the $d$-dimensional square well.  Note that this
analysis says nothing about the off-diagonal terms of $({\Lambda}_{\mu
\nu})_{nn'}$ for a system with degenerate transverse modes; for such
systems, there may indeed be a nonvanishing gauge potential.

\section{Conclusions}

\label{s18}

We have rigorously derived the effective Hamiltonian of a constrained
quantum system by considering the limit as the restoring force becomes
infinite.  In doing so, we have been careful to avoid unnecessary
assumptions on the curvature of the ambient space, the form of the
constraint manifold, and the manner of the constraining potential.
This general approach yields important new terms in the effective
potential ${\sf V}_{ex}$, as outlined in Sects.~\ref{s16} and
\ref{s3}, as well as a gauge potential in the tangential kinetic energy ${\sf K}_\parallel$, as
outlined in Sects.~\ref{s3} and \ref{s13}.  Furthermore, this general
approach allows our theory to be applied to several examples of
physical importance.  These examples include reaction paths for
molecular reaction and scattering problems, twisted quantum
waveguides, the double pendulum, and models of polymers by rigid
constraints.

Perhaps the most important example of a constrained quantum system is
the quantum rigid body.  Though we lack space to include the analysis
here, we have successfully applied our theory to this case.
Physically, we have in mind such systems as semirigid molecules.  If
we assume that the standard Born-Oppenheimer ordering for semirigid
molecules is valid, then our constrained Hamiltonian reproduces
(through the lowest three orders in the Born-Oppenheimer ordering
parameter) the standard results for the rotation-vibration energy
levels of a semirigid molecule.  (See, for example, Papou\v{s}ek and
Aliev~\cite{Papousek82}.)  For such molecules the gauge potential term
in $K_\parallel$ vanishes due to the harmonic form of the constraining
potential.  (We assume a nondegenerate vibrational state; see
Sect.~\ref{s21}.)  For this reason, a more interesting example would
be one in which the standard semirigid analysis breaks down.  This
occurs, for example, in rigid clusters of molecules held together by
van der Waals forces.  For these systems, the gauge potential will not
in general disappear and should have measurable effects on the
rotation-vibration spectrum.  We will pursue these issues in future
publications.

\section{Acknowledgments}
The author wishes to acknowledge Jerry Marsden and Alan Weinstein, who
were instrumental in the initial motivation of this problem.  The
author is also especially grateful to Robert Littlejohn, for many
extended discussions and thoughtful insight, and to Michael M\"{u}ller
for his careful review of the manuscript.  This work was supported by
the Engineering Research Program of the Office of Basic Energy
Sciences at the U. S. Department of Energy under Contract
No. DE-AC03-76SF00098.

\appendix

\section{A Brief Review of Curves in ${\Bbb R}^3$}

\label{s8}

We cite a few important facts about curves in ${\Bbb R}^3$ which we
need in the body of the paper.  For greater depth, see, for example, Spivak
\cite{Spivak79b}.  Consider a curve ${\bf x}(\alpha)$ in ${\Bbb R}^3$.
The parameterization of the curve is given by $\alpha$ which measures
the arclength along the curve.  Hence the tangent vector $\hat{\bf t}
= d {\bf x}/ d \alpha$ is of unit length.  We denote the principal
normal and the binormal by $\hat{\bf n}$ and $\hat{\bf b}$,
respectively.  They are given by

\begin{eqnarray}
\hat{\bf n} 
& = & {d \hat{\bf t}/ d \alpha \over |d \hat{\bf t}/ d \alpha|}, \\
\hat{\bf b}
& = & \hat{\bf t} \bbox{\times} \hat{\bf n}. 
\end{eqnarray}
The vectors $(\hat{\bf t}, \hat{\bf n}, \hat{\bf b})$ form an
orthonormal right-handed frame.  The derivatives of this frame are
given by the famous Serret-Frenet formulas which may be summarized as

\begin{equation}
{d \over d \alpha} 
\left[ 
\begin{array}{c}
\hat{\bf t} \\
\hat{\bf n} \\
\hat{\bf b}
\end{array}
\right]
= 
\left[
\begin{array}{ccc}
0 & \kappa & 0 \\
-\kappa & 0 & \tau \\
0 & -\tau & 0 
\end{array}
\right]
\left[ 
\begin{array}{c}
\hat{\bf t} \\
\hat{\bf n} \\
\hat{\bf b}
\end{array}
\right],
\end{equation}
where $\kappa(\alpha)$ and $\tau(\alpha)$ are called the curvature and
torsion respectively.  The curvature and torsion have units of
reciprocal length.  The reciprocal of $\kappa$ is the radius of
curvature $\rho = \kappa^{-1}$.

\section{The Second Fundamental Form}

\label{s1}

The external curvature of a submanifold ${\cal C}$ embedded in a
manifold ${\cal A}$ is conveniently specified by a rank three tensor
$T$ called the second fundamental form.  Since the second fundamental
form is of critical importance in the body of this paper, we briefly
review a few of its relevant properties.  For greater detail, see
Refs.~\cite{Kobayashi69,ONeill66}.

Throughout this appendix, ${\bf d}$, ${\bf e}$, ${\bf f}$ denote
arbitrary vector fields tangent to ${\cal A}$ and defined over ${\cal C}$;
${\bf w}$, ${\bf x}$, ${\bf y}$, ${\bf z}$ denote vector fields
tangent to ${\cal C}$; and ${\bf v}$ denotes a vector field normal to
${\cal C}$.  The second fundamental form applied to ${\bf e}$ and
${\bf f}$, denoted $T_{\bf e} {\bf f}$, is a vector field defined by

\begin{equation}
T_{\bf e} {\bf f} 
= P_\perp \nabla_{P_\parallel {\bf e}} P_\parallel {\bf f} 
+  P_\parallel \nabla_{P_\parallel {\bf e}} P_\perp {\bf f}, 
\label{r25}
\end{equation}
where $\nabla$ is the Levi-Civita connection on ${\cal A}$ and $P_\parallel$ and $P_\perp$ are respectively the tangent and normal projection
operators of ${\cal C}$
\footnote{Our definition of the second fundamental form differs in the choice of domain and range from that in Ref.~\cite{Kobayashi69}.  We follow the definition of Ref.~\cite{ONeill66}.}.  It is straightforward to verify that $T$
is in fact a tensor.  Furthermore, the second fundamental form
satisfies the identities

\begin{eqnarray}
\langle {\bf d}, T_{\bf e} {\bf f} \rangle 
& = & - \langle {\bf f}, T_{\bf e} {\bf d} \rangle, 
\label{r23} \\
T_{\bf x} {\bf y} & = & T_{\bf y} {\bf x},
\label{r24}
\end{eqnarray}
where $\langle \; ,\; \rangle$ denotes the Riemannian metric on ${\cal
A}$.  In terms of the components $T_{abc} = \langle {\bf E}_a,
T_{{\bf E}_c} {\bf E}_b \rangle$ introduced in Sect.~\ref{s19}, we have

\begin{eqnarray}
T_{abc} & = & - T_{bac}, 
\label{r136} \\
T_{aij} & = & T_{aji}, 
\label{r137} \\
T_{ab\mu} & = & 0, 
\label{r138} \\
T_{\mu \nu a} & = & T_{ija} = 0
\label{r139},
\end{eqnarray}
where the first two equations are simply component forms for
Eqs.~(\ref{r23}) and (\ref{r24}) and the last two follow easily from
Eq.~(\ref{r25}).

In Sects.~\ref{s20} and \ref{s16}, we need the Gauss equation, a
well-known identity relating the second fundamental form $T$, the
Riemannian curvature $\hat{R}$ of ${\cal C}$, and the Riemannian
curvature $R$ of ${\cal A}$.  The Riemannian curvatures are defined by

\begin{eqnarray}
R_{ {\bf d} {\bf e}} {\bf f} 
& = & (\nabla_{\bf d} \nabla_{\bf e} 
- \nabla_{\bf e} \nabla_{\bf d} 
- \nabla_{[ {\bf d}, {\bf e}]} ){\bf f}, 
\label{r31} \\
\hat{R}_{ {\bf x} {\bf y}} {\bf z} 
& = & (\hat{\nabla}_{\bf x} \hat{\nabla}_{\bf y} 
- \hat{\nabla}_{\bf y} \hat{\nabla}_{\bf x} 
- \hat{\nabla}_{[ {\bf x}, {\bf y}]} ){\bf z}, 
\end{eqnarray}
where $\hat{\nabla}$ denotes the Levi-Civita connection on ${\cal C}$.
The Gauss equation is then \cite{Kobayashi69}

\begin{equation}
\langle {\bf w}, R_{{\bf x} {\bf y}} {\bf z} \rangle
= \langle {\bf w}, \hat{R}_{{\bf x} {\bf y}} {\bf z} \rangle
+ \langle T_{\bf x} {\bf z}, T_{\bf y} {\bf w} \rangle
- \langle T_{\bf y} {\bf z}, T_{\bf x} {\bf w} \rangle.
\label{r11}
\end{equation}

\section{The Quantum Kinetic Energy with Respect to a Vielbein}

\label{s9}

We present two expressions for the kinetic energy of a quantum system
on a Riemannian manifold of dimension $n$.  These expressions differ
in the scaling of the quantum wave function.  We refer the reader to
earlier related analyses \cite{Mitchell99,Nauts85,Chapuisat91a} for
derivations and discussion.  

We express the  kinetic energy in terms
of a vielbein.  By a vielbein on a Riemannian manifold, we mean a set
of vector fields ${\bf E}_a$, $a = 1,...,n$, forming a basis of each
tangent space.  The structure constants $\beta_{ab}^c$ of the vielbein are
defined by

\begin{equation}
[ {\bf E}_a, {\bf E}_b ] = \beta^c_{ab} {\bf E}_c,
\end{equation}
where $[\;,\;]$ denotes the Lie bracket. The structure constants
vanish if and only if the vielbein is a coordinate basis, that is if
and only if there exists a set of coordinates $x^a$ such that ${\bf
E}_a =
\partial / \partial x^a$.  We denote the components of the Riemannian
metric with respect to the vielbein by $G_{ab}$ and the inverse of
$G_{ab}$ by $G^{ab}$.

We define the kinetic energy of the quantum system by $K = -\hbar^2
\triangle/2$, where $\triangle$ is the Laplacian.  In terms of the
vielbein, the kinetic energy is \cite{Mitchell99},

\begin{equation}
K 
= {\hbar^2\over 2} {\bf E}_a^\dagger G^{ab} {\bf E}_b
=  {1 \over 2} \pi_a^\dagger G^{ab} \pi_b,
\label{r73}
\end{equation}
where 

\begin{equation}
\pi_a = -i\hbar{\bf E}_a,
\end{equation}
are the momentum operators.  In the above $\dagger$ denotes the
Hermitian conjugate.  In general, the momenta $\pi_a$ are not
Hermitian.  They do, however, satisfy the following useful identity

\begin{equation}
\pi_a^\dagger = \pi_a + \left[\pi_a \ln \sqrt{G}\right] + i \hbar \beta^b_{ab},
\label{r74}
\end{equation}
where $G = \det G_{ab}$.  The bracket notation in Eq.~(\ref{r74})
indicates that the quantity inside the brackets is a scalar; that is,
$\pi_a$ acts only on $\ln \sqrt{G}$.

Often it is useful to scale the original wave function $\varphi$ by
some real positive function $s$ to form a new wave function
$\psi$,

\begin{equation}
\psi = s \varphi.
\label{r83}
\end{equation}
Such a scaling produces a new kinetic energy operator acting on the
new wave function $\psi$.  By conveniently choosing the scale factor
$s$, the new kinetic energy may acquire a more convenient form
than the old kinetic energy.  To demonstrate how the kinetic energy
transforms, we first observe that the scaled wave functions have a
different inner product than the unscaled wave functions.  Denoting
the unscaled inner product by $\langle \; | \; \rangle$, the scaled
inner product $\langle \; | \; \rangle_s$ is defined by

\begin{equation}
\langle \psi | \psi' \rangle_s 
= \left \langle {1 \over s } \psi 
\right. \left| {1 \over s} \psi' \right \rangle,
\label{r140}
\end{equation}
for arbitrary wave functions $\psi$ and $\psi'$.  This scaled inner
product in turn defines a scaled Hermitian conjugate $A^{\dagger(s)}$
of an operator $A$.  Specifically,

\begin{equation}
A^{\dagger(s)} = s^2 A^\dagger {1 \over s^2}.
\label{r75}
\end{equation}
Applying Eq.~(\ref{r75}) to Eq.~(\ref{r74}), we find

\begin{equation}
\pi_a^{\dagger(s)} 
= \pi_a^\dagger - 2[\pi_a \ln s].
\label{r78}
\end{equation}

The scaling of the wave function transforms the kinetic energy
operator $K$ into $K_s = s K (1/s)$.  It can be shown
\cite{Mitchell99} that $K_s$ reduces to

\begin{equation}
K_s 
= \frac{1}{2} \pi_a^{\dagger (s)} G^{ab} \pi_b + V_s,
\label{r76}
\end{equation}
where

\begin{eqnarray}
V_s 
& = & -\frac{1}{2} \left( G^{ab} [ {\pi}_a \ln s] [{\pi}_b \ln s]
+ [ {\pi}_a^{\dagger (s)} G^{ab} [ {\pi}_b \ln s ] ]\right) 
\nonumber \\
& = & \frac{1}{2} \left( G^{ab} [ {\pi}_a \ln s] [ {\pi}_b \ln s]
- [ {\pi}_a^\dagger G^{ab} [ {\pi}_b \ln s ] ]\right). 
\label{r77}
\end{eqnarray}

\begin{figure}
\caption{A twisted quantum waveguide.  The cross sectional shape of
the tube is constant and is chosen to be a triangle with no reflection
symmetry.  (Reflection symmetry would force $\langle \Lambda \rangle$
to vanish.)  The vectors ${\bf E}_1$ and ${\bf E}_2$ determine the
orientation of the sides of the waveguide, and $\alpha$ measures the
distance along the axis.}
\label{f1}
\end{figure}

\end{document}